\documentclass[useAMS,usenatbib,usegraphicx]{mn2e}

\newcommand{\singlecolsize}{0.47}
\newcommand{\doublecolsize}{0.95}
\newcommand{\middlecolsize}{0.80}

\newcommand{\sqdeg}{deg$^2$}
\newcommand{\cosmos}{$(\Omega_{m},\Omega_{\Lambda})_0$}
\newcommand{\Vmax}{$V_{\rm max}$}

\newcommand{\dd}{{\rm d}}
\newcommand{\mass}{{\cal M}}
\newcommand{\Msun}{{\cal M}_{\odot}}
\newcommand{\perMpcsq}{{\rm Mpc}^{-2}}

\newcommand{\sigr}{\sigma_{\rm r}}  \newcommand{\sigb}{\sigma_{\rm b}}
\newcommand{\muer}{\mu_{\rm r}}     \newcommand{\mueb}{\mu_{\rm b}}
\newcommand{\phir}{\phi_{\rm r}}    \newcommand{\phib}{\phi_{\rm b}}

\newcommand{\araa}{ARA\&A}   \newcommand{\aap}{A\&A}
\newcommand{\aj}{AJ}         \newcommand{\apj}{ApJ}
\newcommand{\apjl}{ApJ}      \newcommand{\apjs}{ApJS}
\newcommand{\mnras}{MNRAS}   \newcommand{\nat}{Nature}

\title{Galaxy bimodality versus stellar mass and environment}

\author[I.~K.~Baldry et al.]
{I.~K.~Baldry$^1$, M.~L.~Balogh$^2$, R.~G.~Bower$^3$, 
  K.~Glazebrook$^4$, R.~C.~Nichol$^5$,\newauthor 
  S.~P.~Bamford$^5$, T.~Budavari$^4$\\
$^1$Astrophysics Research Institute, Liverpool John Moores University, 
  Twelve Quays House, Egerton Wharf, Birkenhead CH41\,1LD, UK\\
$^2$Department of Physics and Astronomy, University of Waterloo, 
  Waterloo, ON N2L\,3G1, Canada\\
$^3$Institute for Computational Cosmology, Department of Physics, 
  University of Durham, South Road, Durham DH1\,3LE, UK\\
$^4$Department of Physics and Astronomy, Johns Hopkins University, 
  Baltimore, MD 21218, USA\\
$^5$Institute of Cosmology and Gravitation, University of Portsmouth, 
  Portsmouth PO1\,2EG, UK}

\begin{document}

\date{Accepted 2006 September 15; received 2006 July 28 by MNRAS}

\pagerange{\pageref{firstpage}--\pageref{lastpage}} \pubyear{2006}

\maketitle

\label{firstpage}

\begin{abstract}
  We analyse a $z<0.1$ galaxy sample from the Sloan Digital Sky Survey
  focusing on the variation of the galaxy colour bimodality with stellar mass
  $\mass$ and projected neighbour density $\Sigma$, and on measurements of the
  galaxy stellar mass functions.  The characteristic mass increases with
  environmental density from about $10^{10.6}\Msun$ to $10^{10.9}\Msun$
  (Kroupa IMF, $H_0=70$) for $\Sigma$ in the range 0.1--$10\,\perMpcsq$. The
  galaxy population naturally divides into a red and blue sequence with the
  locus of the sequences in colour-mass and colour-concentration index not
  varying strongly with environment.  The fraction of galaxies on the red
  sequence is determined in bins of 0.2 in $\log\Sigma$ and $\log\mass$ ($12
  \times 13$ bins). The red fraction $f_r$ generally increases continuously in
  both $\Sigma$ and $\mass$ such that there is a unified relation: $f_r =
  F(\Sigma,\mass)$. Two simple functions are proposed which provide good fits
  to the data. These data are compared with analogous quantities in
  semi-analytical models based on the Millennium $N$-body simulation: the
  \citet{bower06} and \citet{croton06} models that incorporate AGN feedback.
  Both models predict a strong dependence of the red fraction on stellar mass
  and environment that is qualitatively similar to the observations.  However,
  a quantitative comparison shows that the \citeauthor{bower06}\ model is a
  significantly better match; this appears to be due to the different
  treatment of feedback in central galaxies.
\end{abstract}


\section{Introduction}
\label{sec:intro}

Galaxies when characterised by their morphology or radial profiles, integrated
or central colours, and total luminosity or stellar mass, exhibit a range of
relationships. These include colour-morphology relations
\citep{Holmberg58,RH94}, and colour-magnitude relations separately for
early-type galaxies \citep{Faber73} and late-type galaxies \citep{CR64}.
While it was often considered that the natural dividing line was between
spirals and ellipticals/lenticulars (e.g.\ \citealt*{TMA82}), it was not until
the multi-wavelength Sloan Digital Sky Survey (SDSS) that the galaxy
population was considered strongly bimodal in colour \citep{strateva01}.  Even
when considering other galaxy properties such as radial profiles, the natural
division is into two galaxy populations \citep{hogg02red,ellis05,ball06}.
Given that large automated imaging surveys are better at defining a galaxy's
colour than morphology, it is more natural to describe a galaxy as being on the
``red sequence'' or ``blue sequence'' rather than being an ``early type'' or
``late type''. This interpretation also has the advantage that galaxy colours
are directly related to the star formation, dust and metal enrichment history
of the galaxy and can thus be more readily interpreted in theoretical models.

A key goal of galaxy evolution theory is to explain the bimodality and the
relationships within each sequence, and there has been considerable work in
this area recently (e.g.\ \citealt{kang05,menci05,menci06,SDH05};
\citealt{AvilaReese06,bower06,cattaneo06,croton06,DB06,perez06}).  The
key ingredient of many of these models is the inclusion of feedback from
active galactic nuclei (AGN). Although AGN feedback (or its equivalent) is
implemented in different ways in each of these models, the overall effect is
to suppress cooling in massive halos.  For example, in \citet{bower06} the AGN
feedback suppresses quasi-hydrostatic cooling flows, while \citet{croton06}
adopt a semi-empirical description for the AGN power related to the Bondi
accretion rate. Bimodality of galaxy colours does not directly result from
these schemes, however, the AGN feedback allows the star formation rate
parameterisation to be adjusted in such a way as to simultaneously obtain a
good description of the colour distribution at faint magnitudes and a good
match to the shape of the luminosity function. In these models red galaxies at
faint magnitudes are predominantly satellite galaxies of brighter systems,
while at bright magnitudes the central galaxies are also red because of the
AGN feedback (e.g.\ fig.~3 of \citeauthor{bower06}).  In the real universe,
however, the association between galaxy colours and their location in the halo
is unlikely to be so simple \citep{weinmann06}, and measurements of the
dependence of galaxy colours on luminosity and redshift are an important
constraint on the new generation of galaxy formation models.  Analysis of the
relationship between galaxy colours and environment will place important
constraints on the processes defining galaxy evolution.

This paper provides a detailed analysis of the variation of the bimodality
with stellar mass and environment.  The plan of the paper is as follows: in
\S\,\ref{sec:earlier-work}, we review a series of papers providing the buildup
to this paper; in \S\,\ref{sec:data}, we describe the data; in
\S\,\ref{sec:results}, we present the results; in \S\,\ref{sec:discussion}, we
discuss the implications and compare the data with models; and in
\S\,\ref{sec:summary}, we summarise the main results.  Fits to the mass
functions are presented in the Appendix.  The data represented in this paper
is available at http://www.astro.livjm.ac.uk/$\sim$ikb/research/ or upon
request.

\subsection{Previous work}
\label{sec:earlier-work}

\citet[hereafter Paper~I]{baldry04} characterised the volume-averaged
colour-magnitude distribution of galaxies by fitting double Gaussian functions
to the colour histograms in magnitude bins. This showed that the red sequence
transitioned from a broader colour distribution at $M_r\sim-19$ to a narrow
distribution for massive early types at $M_r\sim-21$ while the blue sequence
become significantly redder over the range $-20$ to $-22$ (using
centrally-weighted colours). This supported the suggestion of a transition in
galaxy properties around $10^{10.5}\,\Msun$ \citep{kauffmann03B} and the
former is consistent with faint red-sequence galaxies forming more recently
\citep{delucia04,kodama04} but not necessarily in the richest clusters
\citep{Andreon06}. The difference in star formation history for galaxies below
and above the transition mass may be related to the balance between
hydrostatic versus rapid cooling \citep{BD03,keres05,bower06,croton06}, or to
the diversity of star formation histories of satellite galaxies at low masses
\citep{Bower91}.

\citet[hereafter Paper~II]{balogh04} analysed H$\alpha$ emission strength as a
function of galaxy environment. The H$\alpha$ equivalent width (EW)
distribution is bimodal with the distribution of the star forming population
(blue sequence) not depending strongly on environment. The fraction of
galaxies with ${\rm EW}>4$\AA\ varied continuously with environmental
density and there was no evidence of a break density that had been reported
before \citep{lewis02,gomez03}.  This demonstrated the importance of
describing the variation of the bimodal population by comparing the number of
galaxies within each population rather than using a quantity averaged over
both populations.

In Paper~II, the results indicated a dependence of the star-forming fraction
on scales of about 5\,Mpc (after accounting for local environment).  However,
the interpretation is difficult because the small-scale measurement is noisy
and the large-scale measurement could actually be adding information about the
small-scale environment \citep{blanton06}, and \citet{kauffmann04} found no
environmental dependence on scales larger than 1\,Mpc.  It is also known that
for galaxies in clusters properties can depend on environment for scales
smaller than 1\,Mpc \citep{Dressler80,WG91}. These argue for an $N$th nearest
neighbour approach to measuring environmental density as the radius is smaller
in high density regions and expands in low density regions where there are
often insufficient galaxies on small scales (for a fixed luminosity cut).

\citet[hereafter Paper~III]{balogh04bimodal} extended the double Gaussian
fitting of Paper~I to colour histograms across environment and luminosity.  At
fixed luminosity, the mean positions of the sequences become marginally redder
with environmental density. For the red sequence, this can be explained by a
small difference in age between low- and high-density environments of
$\sim2$\,Gyr \citep{thomas05} or less \citep{hogg04,bernardi06}. In contrast,
the fraction of red-sequence galaxies varied strongly with environment as
measured by projected density. No effect related to velocity dispersion of a
group or cluster was detected which is consistent with the morphology-density
relation being similar in groups and clusters \citep{PG84}.

\citet{baldry04conf} confirmed the shifts in colour of the red and blue
sequences, 0.05 and 0.1 in $u-r$, respectively, over a factor of 100 in
projected density. The red fraction varied from 0\% to 70\% for low-luminosity
galaxies and 50\% to 90\% for high-luminosity galaxies (see also
\citealt{tanaka04}).  The effects of environment and luminosity could be
unified in that the fraction of red-sequence galaxies was related to a
combined quantity: $\Sigma_{\rm mod} = (\Sigma/\perMpcsq) \, + \, (L_r /
L_{r,{\rm norm}}) \,$ where $\Sigma$ is the projected density and $L_{r,{\rm
    norm}}$ is luminosity of a galaxy with $M_r=-20.2$. In this paper, this
effect is explored in more detail using a larger sample and by converting
luminosity to stellar mass.

\section{Data}
\label{sec:data}

In this section, the selection of data from the SDSS catalogue and the derived
quantities for each galaxy are described. The basics of the SDSS are described
in \S\,\ref{sec:sdss} while the sample selection is described in
\S\,\ref{sec:samples}. The primary sample consists of spectroscopically
observed galaxies but a larger sample was used to determine spectroscopic
completeness (\S\,\ref{sec:completeness}) and to measure uncertainties in
environmental densities (\S\,\ref{sec:densities}) for which photometric
redshifts were determined (\S\,\ref{sec:photoz}).  Number density corrections
are described in \S\,\ref{sec:volume-density-corr}. The derived quantities
include absolute magnitude $M_r$, rest-frame colour $C_{ur}$, inverse
concentration index ${\cal C}$ (\S\,\ref{sec:absmag-colour-concentration}),
projected density of neighbouring galaxies $\Sigma$ (\S\,\ref{sec:densities}),
and an approximate stellar mass $\mass$ (\S\,\ref{sec:stellar-mass}).

\subsection{Sloan Digital Sky Survey}
\label{sec:sdss}

\begin{figure*}
  \includegraphics[width=\doublecolsize\textwidth]{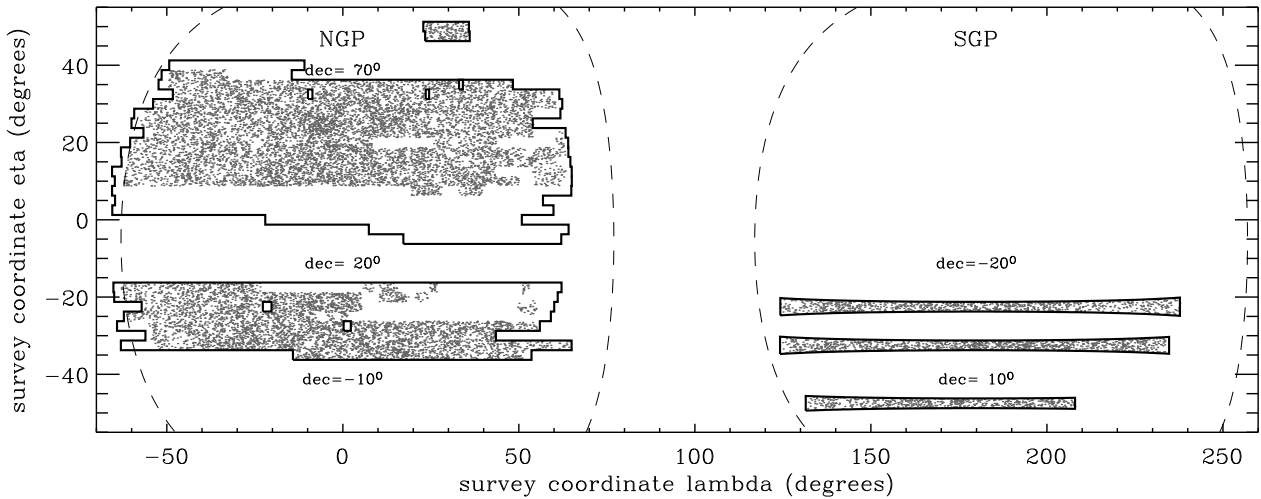}
\caption{Sky coverage for SDSS DR4. 
  The solid lines outline the primary photometric area, the dashed lines
  outline the galactic poles at 20\degr\ latitude, and the points show the
  positions for 4\% of the galaxies that have been observed spectroscopically.
  The edges are are easier to define in SDSS survey coordinates, which is a
  spherical coordinate system [(eta,lambda)=(0,0) corresponds to
  (ra,dec)=(185,32.5)]. The photometric area covers 6670\,\sqdeg\ and the
  spectroscopic area covers 4780\,\sqdeg\ \citep{adelman-mccarthy06}.}
\label{fig:edges}
\end{figure*}

The SDSS is a project, with a dedicated 2.5-m telescope, designed to image
$10^4$\,\sqdeg\ and obtain spectra of $10^6$ objects
\citep{york00,stoughton02}. The imaging covers five broadbands, $ugriz$ with
effective wavelengths of 355, 467, 616, 747 and 892\,nm, using a mosaic CCD
camera \citep{gunn98}.  Spectra are obtained using a 640-fibre fed
spectrograph with a wavelength range of 380 to 920\,nm and a resolution of
$\lambda/\Delta\lambda\sim1800$.

The images are reduced using a pipeline {\tt photo} that measures the
observing conditions, and detects and measures objects. In particular, {\tt
  photo} produces various types of magnitude measurement including: (i)
`Petrosian', the summed flux in an aperture that depends on the
surface-brightness profile of the object; (ii) `model', a fit to the flux
using the best fit of a de-Vaucouleurs and an exponential profile; (iii)
`PSF', a fit using the local point-spread function.  The magnitudes are
extinction-corrected using the dust maps of \citet*{SFD98}.  Details of the
imaging pipelines are given by \citet{stoughton02}.

Once a sufficiently large area of sky has been imaged, the data are analysed
using `targeting' software routines that determine the objects to be observed
spectroscopically. The main sets of targets are: (i) the main galaxy sample
(MGS), an $r<17.77$ selection \citep{strauss02}; (ii) luminous red galaxies
(LRGs), an $r<19.5$ and colour-selected galaxy sample \citep{eisenstein01}; and
(iii) quasi-stellar objects (QSOs; \citealt{richards02}).  The targets from
all the samples are then assigned to plates, each with 640 fibres, using a
tiling algorithm \citep{blanton03tile}.  The main restriction is that two
fibres cannot be placed within $55''$ on the same plate.

Spectra are taken using, typically, three 15-minute exposures in moderate
conditions (the best conditions are used for imaging).  The signal-to-noise
ratio (S/N) is typically 10 per pixel ($\approx1$--2\AA) for galaxies in the
MGS. The pipeline {\tt spec2d} extracts, and flux and wavelength calibrates,
the spectra. The spectra are then analysed by another pipeline that classifies
and determines the redshift of the object. The redshift success is about 99\%
for galaxies considered in this paper.

\subsection{Sample selection}
\label{sec:samples}

The data used in this paper are from the SDSS Data Release Four (DR4) 
\citep{adelman-mccarthy06}. The primary sample consists of 151\,642 galaxies
with $r<17.77$ in the redshift range 0.010--0.085 (Sample~C), with
environmental densities determined from the distance to the 4th and 5th
nearest neighbours with $M_r<-20$.  However, the spectroscopic sample is not
100\% complete; some galaxies are missed because of fiber-placement
restrictions or are not yet observed, and there are edges to the sky coverage.
Figure~\ref{fig:edges} shows the sky coverage for the photometric and
spectroscopic data. In order to accurately account for missed spectra and to
calculate completeness corrections, a larger photometric sample was
considered.  The galaxy sample selection and calculation of environmental
densities are outlined below.

Objects were selected from the `CasJobs' catalogue archive 
server\footnote{http://casjobs.sdss.org/CasJobs/} with criteria
$
   10.0 <  r < 18.0 \: ,
$
where $r$ is the Petrosian magnitude corrected for Milky-Way extinction, and 
$
    r_{\rm PSF} - r_{\rm model} > 0.25 \: .
$
Photometric data (`Galaxy' table) and spectroscopic data (`SpecObjAll' table)
where available were extracted. This produced an initial sample of 
1\,016\,565 objects after removing duplicates (Sample~A).

\begin{figure}
  \includegraphics[width=\singlecolsize\textwidth]{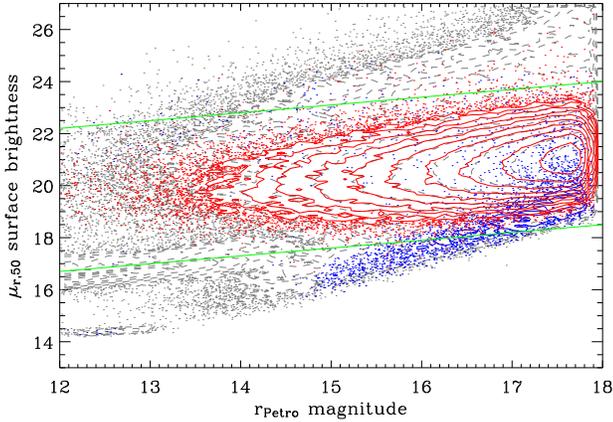}
\caption{Surface brightness versus magnitude. 
  The grey points and dashed contours represent all objects in Sample~A; the
  red points and contours represent spectroscopically confirmed galaxies; and
  the blue points and contours represent stellar systems. The green lines show
  the limits imposed for photometric objects selected for Sample~B. Note that
  most stars are rejected by the `not saturated' criteria rather than the
  lower limit shown here.}
\label{fig:sb-mag}
\end{figure}

The next stage was to remove low-surface brightness artifacts and stars. A 
modified surface brightness was defined:
$
        \mu_{\rm mod} = \mu_{r,{\rm 50}} - 0.3(r-18)
$
where $\mu_{r,{\rm 50}}$ is the mean surface brightness (SB) within the
Petrosian half-light radius (eq.~5 of \citealt{strauss02}). The slope of 0.3
per magnitude was determined empirically from the SB-magnitude distribution so
that the modified quantity was a better separator for stars, galaxies and
artifacts. 
Figure~\ref{fig:sb-mag} shows SB versus magnitude for various samples and the
dividing lines.  Objects with no spectroscopy were then trimmed to
$
    18.5 < \mu_{\rm mod} < 24.0
$
and `not saturated' while spectroscopic objects were selected with 
$
    0.005 < z < 0.3 .
$
The trimmed sample contained 850\,906 galaxies (Sample~B). 

\subsection{Spectroscopic completeness}
\label{sec:completeness}

\begin{figure}
  \includegraphics[width=\singlecolsize\textwidth]{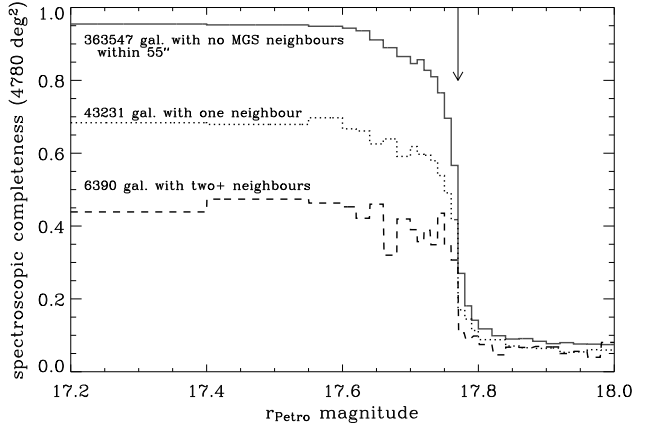}
\caption{Spectroscopic completeness versus magnitude.
  The three lines represent galaxies with 0, 1 or $\ge2$ neighbours within
  $55''$ (corresponding to the `fibre collision' radius).  The arrow shows the
  limit for MGS selection; the completeness drops prior to that because of
  changes in catalogue photometry and targeting algorithms. Here we have used
  the DR4 best photometry.}
\label{fig:completeness}
\end{figure}

Spectroscopic completeness was determined from a related sample that included
all spectra and was reduced to spectroscopic regions (634\,395 galaxies,
478\,385 with $r<17.77$). Galaxies were divided into classes based on their
magnitude and the number of $r<17.77$ neighbours within $55''$ (0, 1 or
$\ge2$).  The completeness was given by the number with spectra divided by the
total number in a class.
Figure~\ref{fig:completeness} shows completeness versus magnitude. As expected
the completeness is low above 17.77 and it becomes colour dependent (LRG
selection). Thus, the final sample (Sample~C) is restricted to this MGS limit
where the completeness corrections are reliable.  It is also clear that in
order to determine the relative numbers of galaxies in different environments,
it is important to account for the change in completeness as a function of the
number of neighbours.

\subsection{Photometric redshifts}
\label{sec:photoz}

Photometric redshifts were determined for Sample~B using $\chi^2$ fitting to
`training set' galaxies with spectroscopic redshifts.  The training set
consisted of 396\,302 galaxies out of a possible 410\,206, with some galaxies
rejected because of low redshift confidence, or extreme colours or colour
errors.  The data fitted for each galaxy were $r_{\rm Petro}$ and 4 `model'
colours: $u-g$, $g-r$, $r-i$, $i-z$.  Thus, the underlying assumption is that
the redshift-magnitude-colour distribution is similar for the galaxies with
missing spectra as for the training set.  Softening errors were added in
quadrature, prior to determination of the $\chi^2$ values.  These were 0.1 for
the magnitude (larger for $r<13$) and 0.02 for the colours.\footnote{The
  $\chi^2$ values were determined only using the errors for the galaxy being
  considered, not the training-set galaxies, otherwise higher weight would be
  given to training-set galaxies with larger errors! For testing and
  calibration of the redshift errors, training-set galaxies were not matched
  to themselves.} The photometric redshift is then given by the weighted mean
of training-set redshifts that fitted within $\chi^2<15$ (99\% limit for 5
degrees of freedom).  The weight for each match is given by
$\exp({-\chi_i^2/2})\,w(z_i)$, where $w(z_i)$ is a minor adjustment to reduce
the implicit weight given to redshift peaks within the training set.  Objects
with no matches within the limit or with few matches in the range
$10<\chi^2<15$ were assumed to be outside the redshift range of the training
set (i.e.\ stars, or $z>0.3$) and not considered in further analysis (13\,741
objects).  An initial photometric redshift error was determined from the
standard deviation (with weights). These errors were then multiplied by a
factor to give 95\% confidence-limit errors:
\begin{equation}
  z_{\rm err, 95} =  f_{95} 
  \left( \frac{|z_{\rm photo} - z_{\rm spec}|}{z_{\rm err, initial}} \right) 
  z_{\rm err, initial}
\label{eqn:photoz-error}
\end{equation}
where $f_{95}()$ is the 95-th percentile of the expression determined for
galaxies with spectroscopic redshifts.  After calibration, the median value of
$z_{\rm err, 95}$ is 0.04 with typical values from 0.02 to 0.07. A minimum
error of 0.02 was used in subsequent analysis.

\subsection{Absolute magnitude, rest-frame colour and concentration index}
\label{sec:absmag-colour-concentration}

The $r$-band absolute magnitude used in this paper is given by
\begin{equation}
  M_r = r - k_r - 5\log(D_L/{\rm \,10\,pc})
\label{eqn:abs-mag}
\end{equation}
where: $r$ is the Milky-Way-extinction-corrected Petrosian magnitude;
$D_L$ is the luminosity distance for a cosmology with \cosmos\ =
(0.3,0.7) and $H_0 = {\rm\,70\,km\,s^{-1}\,Mpc^{-1}}$, and;
$k_r$ is the k-correction using the method of
\citet{blanton03kcorr};\footnote{The $k$-corrections were derived from
  {\tt kcorrect} v.~4.1.3.}
and rest-frame colour considered is given by
\begin{equation}
  C_{ur} = (u_{\rm model} - k_u) - (r_{\rm model} - k_r) \: .
\label{eqn:colour}
\end{equation}
This was chosen as the optimum bimodality divider (Paper~I).  In effect,
`model' colours are centrally-weighted colours; the profile used is the best
fit defined in the $r$-band light.

Galaxies with photometric redshifts have a range of uncertainty in their
absolute magnitude. For calculation of environmental densities, their
$k$-corrections were approximated by
$
k_r = [ 0.3 + 0.3 (u-r)_{\rm model} ] z_{\rm test}
$
where $z_{\rm test}$ takes different values: the spectroscopic redshifts of
the galaxies for which environmental density is measured.

The (inverse) concentration index was used in some of our analysis.
The is given by 
$
{\cal C} = R50 / R90
$
where $R50$ and $R90$ are the radii containing 50\% and 90\% of the Petrosian
flux, averaged for the $r$ and $i$ bands. For typical galaxies, ${\cal C}$
ranges from 0.3 (concentrated) to 0.55; for comparison, a uniform disk would
have ${\cal C}=0.75$.

\subsection{Environmental densities}
\label{sec:densities}

Densities were determined for Sample~C using the information from Sample~B.
For environmental densities around each galaxy, we determine
\begin{equation}
  \Sigma_N = \frac{N}{\pi d_N^2}
\label{eqn:sigma_n}
\end{equation}
where $d_N$ is the projected comoving distance to the $N$th nearest neighbour
that is a member of the density defining population (DDP) and that is within
the allowed redshift range.  The redshift range is $\pm \Delta z c =
1000{\rm\,km/s}$ for neighbouring galaxies with spectroscopic redshifts and
$\pm z_{err,95}$ for those with photometric redshifts only. The DDP are
galaxies with $M_r < M_{r,{\rm limit}}-Q(z-z_0)$ where $M_{r,{\rm limit}} =
-20$, $Q=1.6$ is the evolution determined by \citet{blanton03ld} and
$z_0=0.05$. For galaxies with photometric redshifts only, $M_r$ was determined
as if the galaxy were at the redshift of the galaxy whose environment was
being measured. If the distance to the photometric edge $d_{\rm edge}$ were
less than $d_N$ then $\pi d_N^2$ in Eq.~\ref{eqn:sigma_n} was replaced by the
area within a chord crossing the circle, i.e., assuming the edge was straight.

The best estimate $\Sigma$ was obtained by averaging $\log \Sigma_N$ for $N=4$
and 5, and by averaging the calculation with spectro-$z$ only and the full
sample~B.  In addition $\Sigma_{\rm min}$ was determined using $d_N$ with
spectro-$z$ only, and $\Sigma_{\rm max}$ using the smaller of $d_N$ with
Sample~B and $d_{\rm edge}$. The uncertainty in $\log\Sigma$ was given by the
larger of $\log\Sigma_{\rm max} - \log\Sigma$ and $\log\Sigma -
\log\Sigma_{\rm min}$.  For various environmental analyses, galaxies were only
included if the uncertainty were less than 0.4\,dex or less than 1 bin for
galaxies in the minimum or maximum density bins. This more complicated
procedure than simply using spectro-$z$ only and discarding galaxies near an
edge ensures that the maximum number of galaxies can be used in any analysis;
and it accurately accounts for `fibre collisions'.

Variations in the above parameters ($\Delta z c = 1000{\rm\,km/s}$, $M_{r,{\rm
    limit}}=-20$, $N=4$ and 5) used to determine $\Sigma$ were tested.  The
results were not strongly dependent on the exact values, and the values chosen
were near optimum and consistent with earlier papers (Paper~II and Paper~III).
For typical galaxies, $\Sigma$ ranges from $0.05\,\perMpcsq$ (void-like
regions) to $20\,\perMpcsq$ (clusters); for comparison,
$\Sigma=0.5\,\perMpcsq$ for the median environment around DDP galaxies whereas
$\Sigma$ would be $0.14\,\perMpcsq$ if DDP galaxies were distributed
uniformly.

\begin{figure}
  \includegraphics[width=\singlecolsize\textwidth]{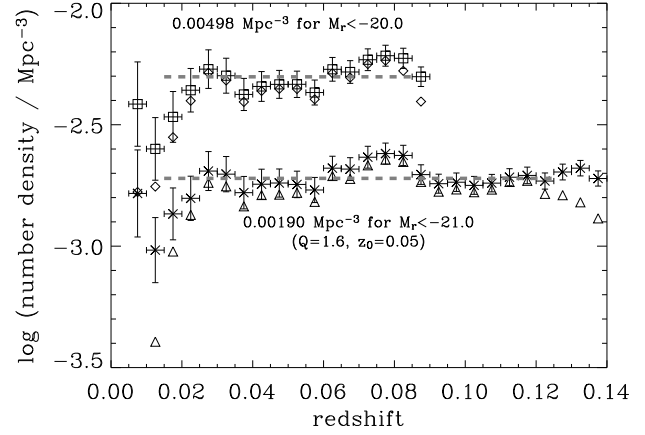}
\caption{The number densities of galaxies versus redshift.
  The squares with error bars (including photometric redshifts) and diamonds
  (spectroscopic redshifts only) represent the number densities of galaxies
  more luminous than $-20$ (evolution corrected to $z=0.05$ with $Q=1.6$,
  \citealt{blanton03ld}); this is the density defining population used in this
  paper.  The crosses with error bars and triangles represent galaxies more
  luminous than $-21$.  Above about $z>0.09$ and $z>0.13$, the counts become
  incomplete in the $r<18$ sample for the $M_r<-20$ and $M_r<-21$ populations,
  respectively. At $z<0.02$ there is incompleteness because of problems
  with automatic determination of photometric properties.  Using $Q=1.6$ for
  the luminosity evolution, the number density evolution is near flat across
  the reliable redshift range, shown by the dashed lines.}
\label{fig:ddp-densities}
\end{figure}

Figure~\ref{fig:ddp-densities} shows the number density of the DDP population
versus redshift and the density of a more luminous $M_r<-21$ population.  The
near constant number density up to the completeness limit justifies the use of
the evolution parameter determined by \citet{blanton03ld}. Sample~C was 
restricted to the redshift range 0.010 to 0.085 for reliability of the 
environmental density measurements. 

\subsection{Volume and density corrections}
\label{sec:volume-density-corr}

The DDP is volume limited but we also consider galaxies fainter than
$M_r=-20$. For these and for some galaxies near $-20$, a volume correction is
applied to the number density using a \Vmax\ method \citep{Felten76}.  Here,
the weight of a galaxy for computing number densities is given by a factor
$1/V_{\rm max}$ where \Vmax\ is the maximum volume over which the galaxy could
be observed within the $r<17.77$ limit and the survey volume ($0.010<z<0.085$
over 1.52 steradians; $2.3\times10^7{\rm\,Mpc}^3$).

The added complication when considering environment is the fact that the
contribution of a particular environment bin to the volume-averaged number
density can vary significantly as a function of redshift.  To correct for
this, we determine a relative volume ($f_V$) that a galaxy in a particular
environment could be observed over the redshift range given by \Vmax\ 
($0.010<z<z_{\rm max}$). This relative volume is a function of $z_{\rm max}$
and $\Sigma$ and can be determined by measuring the number densities of bright
galaxies that can be viewed over the whole survey volume.  The relative volume
was determined in bins of 0.05 in $z$ and in $\Sigma$ (depending on the bins
being used).  In other words, $f_V$ is the number density of bright galaxies
with $z<z_{\rm max}$ relative to the number density covering the whole
redshift range (with reduced survey area to avoid edge effects).  The number
density of galaxies with a given $z_{\rm max}$ and $\Sigma$ was renormalised
by $1/f_V$. This factor $f_V$ is in the range 0.6--1.1 for $z_{\rm
  max}\ga0.03$ over 12 environment bins; $f_V$ is on average less than unity
because galaxies are rejected when the uncertainty in $\Sigma$ is too large.

To summarise, there are three number density corrections for each galaxy in
Sample~C: (i) $1/C_s$ where $C_s$ is the spectroscopic completeness which is a
function of $r$ and the number of MGS neighbouring galaxies within $55''$
(\S\,\ref{sec:completeness}); (ii) $1/V_{\rm max}$ where \Vmax\ is a function
of the absolute magnitude and the galaxy's colours which determine
$k$-corrections; (iii) $1/f_V$ where $f_V$ is the relative volume which is a
function of $z_{\rm max}$ and $\Sigma$ when dividing the population into
environment bins.

\subsection{Conversion to stellar mass}
\label{sec:stellar-mass}

\begin{figure}
  \includegraphics[width=\singlecolsize\textwidth]{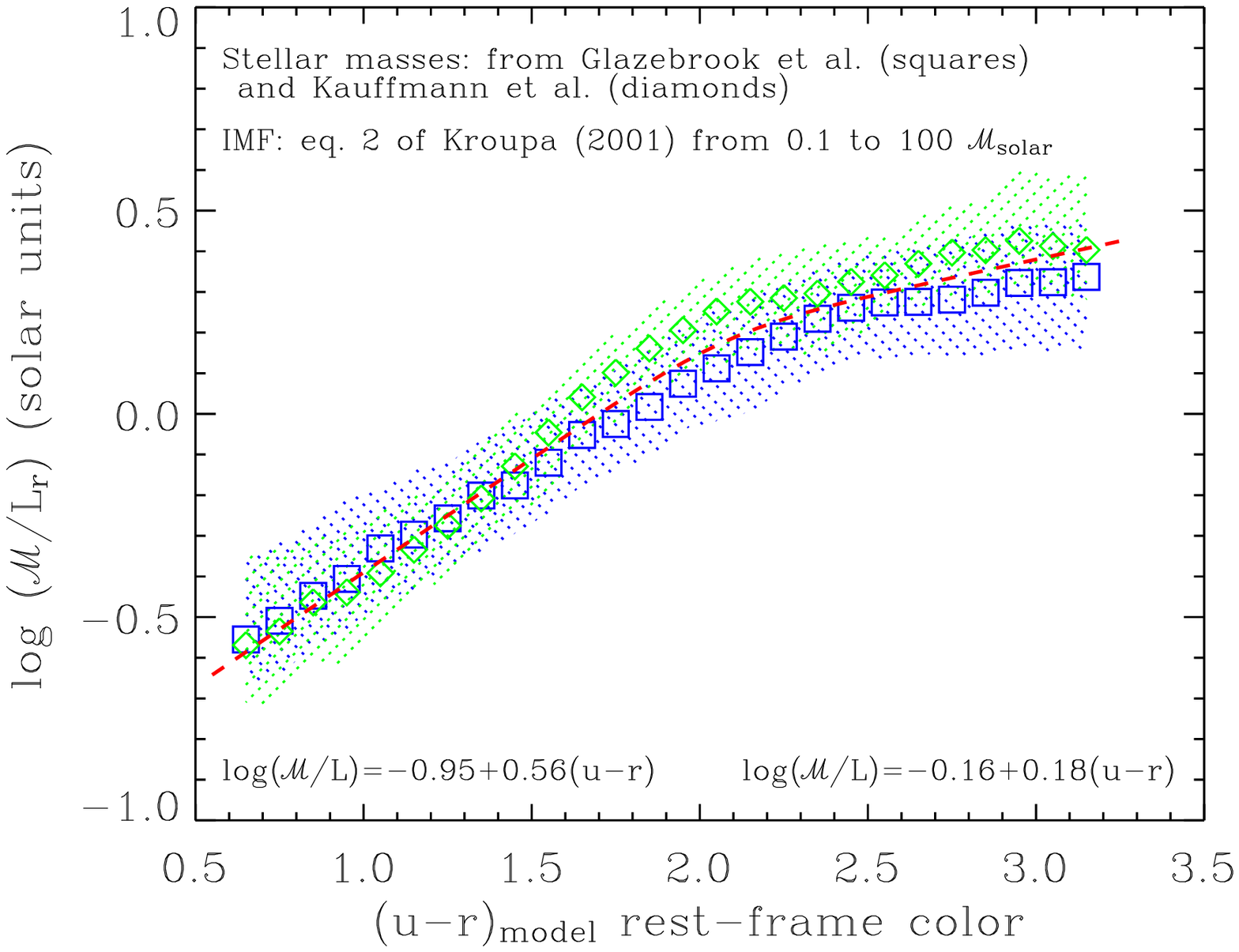}
\caption{Correlation of galaxy stellar mass-to-light ratio with colour.
  Stellar masses were determined for a sample of $10^4$ galaxies.  The blue
  squares and dotted-line region represent the median in 0.1 color bins and
  1-sigma ranges computed by K.G.\ \citep{glazebrook04}. The green diamonds
  and region represent stellar masses computed by the MPA Garching group
  \citep{kauffmann03A}. The red dashed line represents a broken line fit to
  the data, with the coefficients shown in the plot.}
\label{fig:ml-vs-u-r}
\end{figure}

The use of stellar mass as a galaxy variable has been widely promoted (e.g.\ 
\citealt{TG88,PS96,BE00,BFS02,bell03,perez-gonzalez03,conselice05,
  gallazzi05,pannella06}).  The total stellar mass of a galaxy is a more
fundamental quantity than luminosity as has been shown by, for example,
\citet{kauffmann03A}.  Galaxy stellar masses are typically derived by fitting
stellar-population synthesis models to colours or spectra.  However, there are
systematic uncertainties that depend on the underlying set of allowed
star-formation histories in the models as well as uncertainties in the models.
Here, we use an approach similar to that of \citet{BdJ01} whereby we use a
stellar mass-to-light ratio that is a function of one colour only.  While
uncertainties will remain, the method has the advantage of retaining a simple
relation between the derived physical quantity, the selection function, and
the rest-frame colour used in our analysis.  Figure~\ref{fig:ml-vs-u-r} shows
a plot of stellar mass-to-light ratio versus $C_{ur}$ colour for $10^4$
galaxies analysed by the MPA Garching group and by K.G. The dashed line
represents the relation that we use in our analysis.\footnote{We use $\mass$
  for mass and $M$ for absolute magnitudes.  The conversion is given by $\log
  (\mass / \Msun) = (M_{r\odot} - M_r)/2.5 + \log (\mass/L_r)$ where
  $M_{r\odot} = 4.62$ \citep{blanton01}.  The computation of stellar masses by
  the MPA Garching group uses the spectral features, D4000 and H$\delta$, and
  the Petrosian $z$-band magnitude and is described by \citet{kauffmann03A};
  while the method of K.G.\ is described by \citet{glazebrook04} except
  Petrosian magnitudes were used here. For both analyses, $\mass/L_r$ ratios
  were obtained using the \citet{Kroupa01} IMF (eq.~2 of that paper, from
  $0.1\Msun$ to 100 or $120\Msun$).}

\section{Results}
\label{sec:results}

\subsection{Colour-concentration relations as a function of stellar mass}
\label{sec:colour-conc}

\begin{figure*}
  \includegraphics[width=\doublecolsize\textwidth]{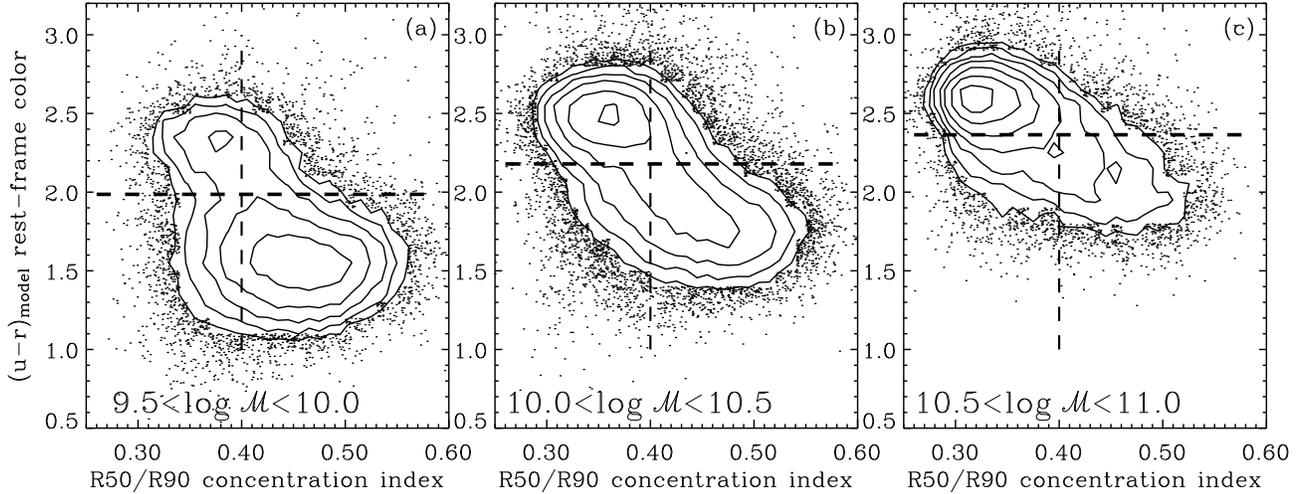}
\caption{Colour versus concentration index for different stellar-mass ranges.
  The points and solid contours represent galaxies in our sample.  The
  contours are logarithmically spaced in number density with 4 contours per
  factor of 10.  The dashed lines show the position of possible colour or
  concentration-index dividers.}
\label{fig:colour-conc}
\end{figure*}

In Paper~I and Paper~III, we divided the galaxy population by using
double Gaussian fitting to the colour functions in luminosity bins.  This
separates the galaxy population into red and blue sequences.  There was no
consideration of the morphology or structure of the galaxies.  This method is
accurate if the colour functions are truly the sum of Gaussian distributions
and if the errors are small. A more robust method of outlining the two
sequences is to consider a joint distribution in colour and structure
\citep{driver06}. 
Figure~\ref{fig:colour-conc} shows the distribution of observed galaxies in
colour versus concentration index for three different stellar mass ranges.

The peak of the red sequence gets redder (2.3 to 2.6) and more concentrated
(0.38 to 0.32) with increasing mass ($\log\mass$ from 9.5 to 11.0); while
the blue sequence gets redder (1.5 to 2.1) with approximately the same
concentration (0.45). This means that the natural dividing lines of the
population are around ${\cal C}=0.4$ for all the stellar masses (vertical
dashed lines in the figure) but the dividing line in colour gets redder
(horizontal dashed lines). At lower masses, dividing using colour is
significantly better than dividing using concentration index. At higher
masses, while colour is still a reasonable divider, concentration index can be
useful in robustly determining the colour mean and dispersion of each sequence.

In Paper~I, a method for fitting double Gaussians to the colour functions is
described. At each luminosity or mass bin, there are six parameters,
amplitude, mean and dispersion for each sequence, given by
$\phir,\muer,\sigr,\phib,\mueb,\sigb$. When the counts are low or the
sequences are significantly merged together in colour (e.g.\ at the high-mass
end), the solution is not well defined. With the robust method, $\muer$ and
$\sigr$ are determined using galaxies with predominately high concentration
(full weight for ${\cal C} < 0.4$) while $\mueb$ and $\sigb$ are determined
using galaxies with low concentration (${\cal C} > 0.4$). The amplitudes of
each Gaussian are still determined using a fit to both populations but with
the $\mu$ and $\sigma$ values fixed.

This method of determining the positions of the sequences is used in
\S\,\ref{sec:colour-mag-mass}, which then determines the best-fit dividing
line in color, and in \S\,\ref{sec:enviro-colour-mass-conc}. In
\S\,\ref{sec:enviro-mass-unify}, only the dividing line is used.

\subsection{Volume-averaged colour-magnitude and colour-mass relations}
\label{sec:colour-mag-mass}

\begin{figure*}
  \includegraphics[width=\doublecolsize\textwidth]{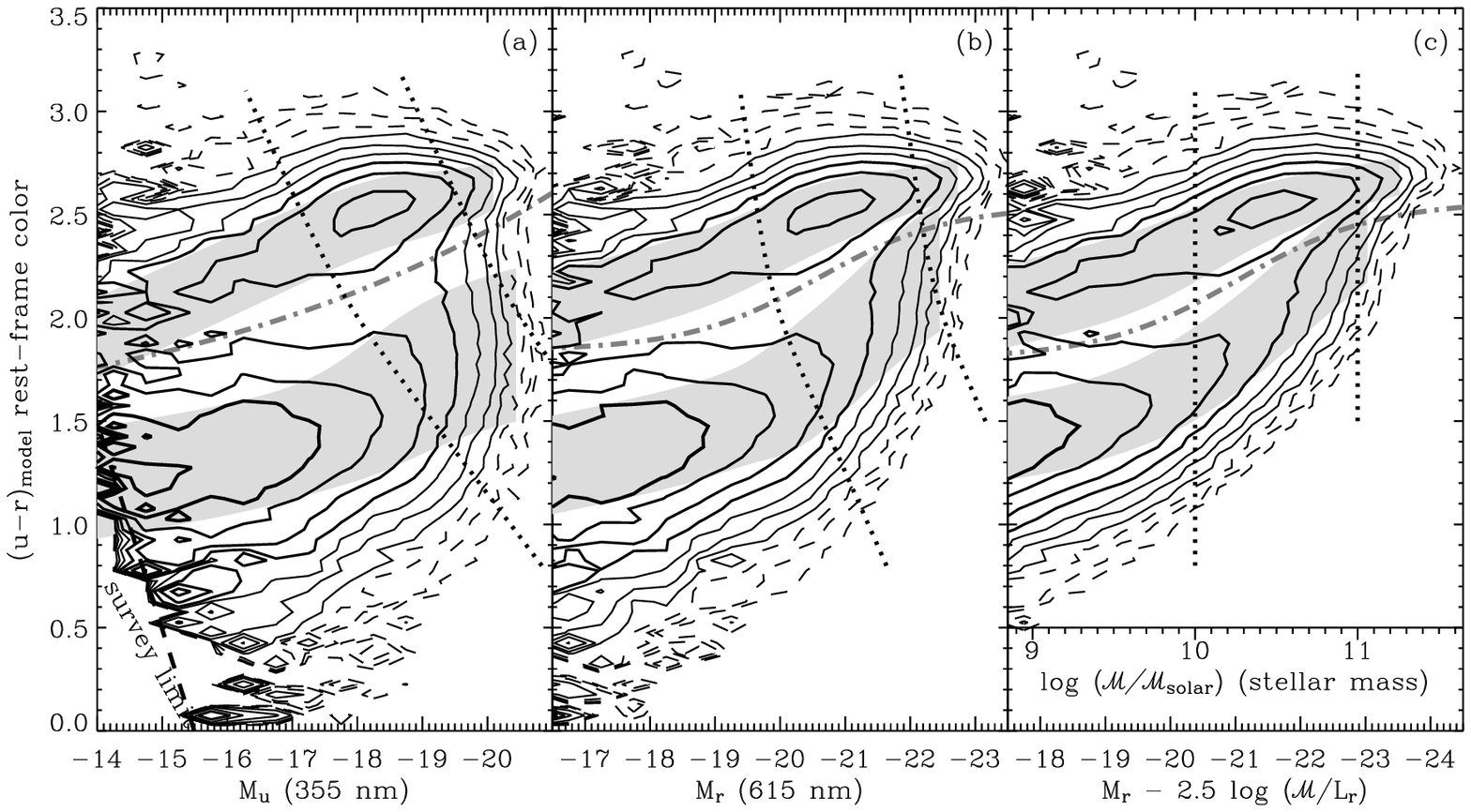}
\caption{Comparison between galaxy colour-magnitude and colour-mass relations. 
  Colour is plotted versus (a) $M_u$, (b) $M_r$ and (c) $\log\mass$. The dashed
  and solid lines represent logarithmically-spaced density contours with 4
  contours per factor of 10; the number densities are completeness and volume
  corrected.  The dotted lines represent galaxies with stellar masses of
  $10^{10}$ and $10^{11}\,\Msun$. The grey regions represent the colour means
  and $\pm1$-sigma ranges of the red and blue sequences. The dash-and-dotted
  lines represent the best-fit dividers between the sequences based on the
  criteria of Paper~I.}
\label{fig:compare-cm}
\vspace*{0.5cm}
  \includegraphics[width=\doublecolsize\textwidth]{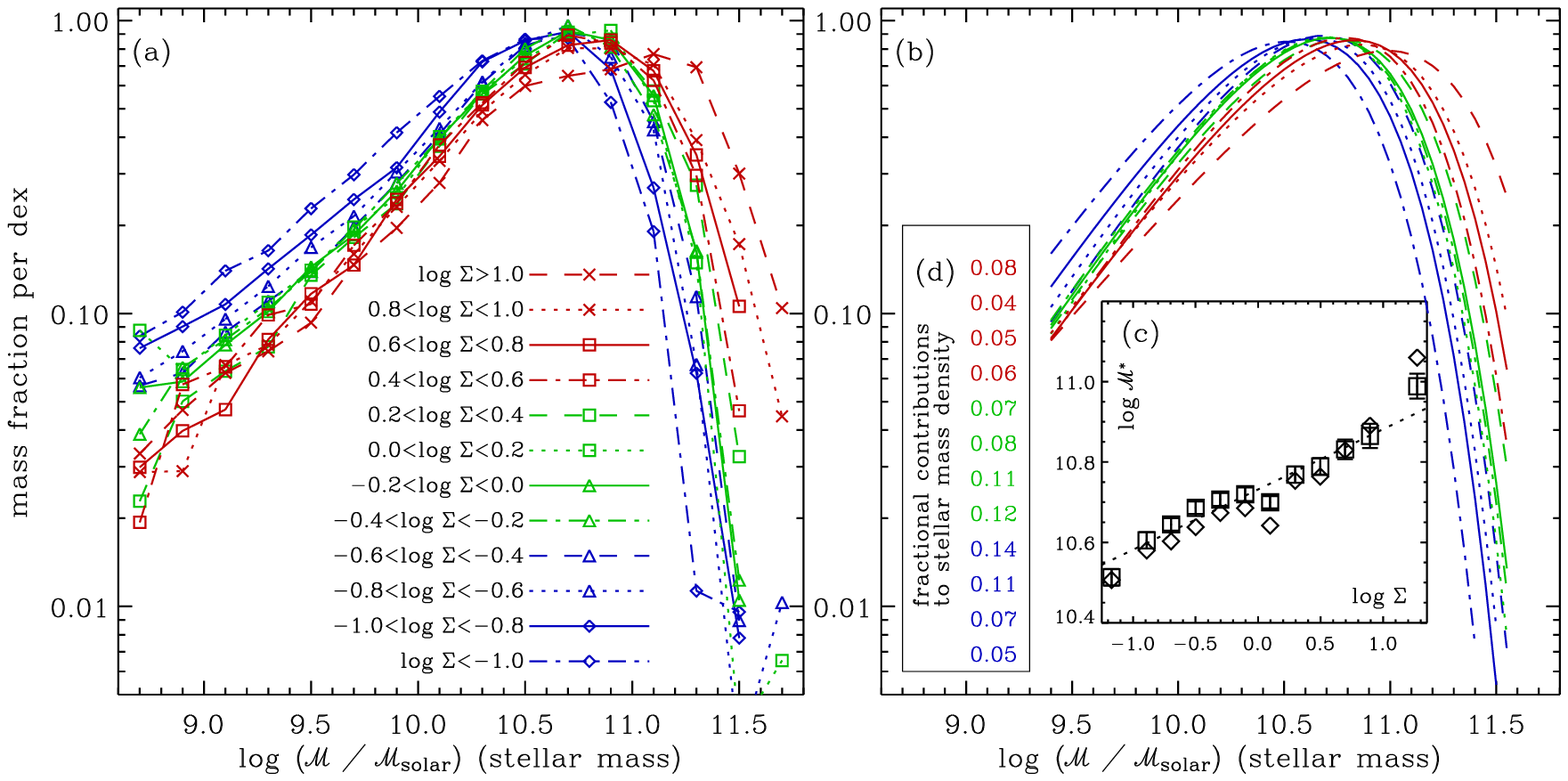}
\caption{Galaxy stellar mass functions.
  In panels~(a) and~(b), the symbols and lines represent different
  environments as shown in the legend of panel (a). Note the unusual $y$-axis,
  with respect to standard mass or luminosity functions, shows clearly that
  the `peak' in the mass function is near the Schechter break
  ($\log\mass^{*}$).  Panel~(a) shows the binned functions (the median errors
  are 5\%) while panel~(b) shows the Schechter function fits. Inset panel~(c)
  shows $\log\mass^{*}$ versus $\log \Sigma$. The diamonds represent
  $\log\mass^{*}$ as defined by the parameter in the fit while the squares
  represent the peaks of the fits given by $\log[\mass^{*}(\alpha+2)]\,$.
  Note that a standard Schechter function fit is only valid over the range
  from about 9.5 to 11.5, and there is a statistically significant upturn in
  the faint-end slope at low masses (cf.\ Paper~I, \citealt{blanton05},
  Appendix).  Inset panel~(d) shows the approximate fractional contribution to
  the stellar mass density of the universe ($\Omega_{\rm stars}\approx0.002$)
  for each environmental bin from high to low densities (top to bottom). Plots
  of the GSMFs using double Schechter fits and using number versus $\log\mass$
  are shown in Fig.~\ref{fig:mass-functions2}.}
\label{fig:mass-functions}
\end{figure*}

Figure~\ref{fig:compare-cm} shows a comparison between colour-magnitude
and colour-mass relations: $C_{ur}$ versus (a) $u$-band absolute magnitude, (b)
$r$-band absolute magnitude, and (c) logarithm of the stellar mass.  Note how
the contours are almost vertical at the luminous end in
Fig.~\ref{fig:compare-cm}(a) whereas the red sequence is dominant at the
high mass end in Fig.~\ref{fig:compare-cm}(c).

In order to fit the mean positions of the sequences, we used a tanh plus a
straight-line function as per Paper~I.  A general form of this `${\cal T}$
function' is given by
\begin{equation}
  {\cal T}(x) = 
  p_0 + p_1 x + q_0 \tanh\left[ \frac{x - q_1}{q_2} \right]
\label{eqn:sl-tanh}
\end{equation}
where $p_{0,1}$ are the straight line parameters; and $q_{0,1,2}$ are the tanh
parameters.  The fitting of the sequences followed the iteration outlined in
\S\,4.2 of Paper~I except with the addition of the robust method outlined in
\S\,\ref{sec:colour-conc} above for the double Gaussian fitting.

The grey regions in Fig.~\ref{fig:compare-cm} show the mean position and
dispersion along each sequence.  A key point is that the blue sequence is
significantly narrower versus stellar mass than luminosity, particularly
near-UV luminosity (the color dispersion is about 0.23, 0.31, 0.37 for
blue-sequence galaxies with $\log\mass\sim10.5$, $M_r\sim-21.0$,
$M_u\sim-19.5$, respectively). This demonstrates that stellar mass is closer
to a fundamental predictor of a galaxy's colour than luminosity.  The
dash-and-dotted lines in the figure represent the best-fit dividing lines
between the red and blue sequences using the optimal criteria of Paper~I.

When dividing the galaxy population by environment, it becomes more difficult
to proceed with the full fitting procedure. For cases where many environmental
bins are used, the number of galaxies in each sequence is defined by galaxies
redder and bluer than the best-fit dividing line [dash-and-dotted line in
Fig.~\ref{fig:compare-cm}(c)]: $C_{ur,{\rm divide}}$ varies from 1.8 to
2.5 for $\log\mass$ from 9.0 to 11.5; ${\cal T}$ function with parameters of 
$p_{0,1}=(2.18,0)$, $q_{0,1,2}=(0.38, 10.26, 0.85)$.

\begin{figure*}
  \includegraphics[width=\doublecolsize\textwidth]{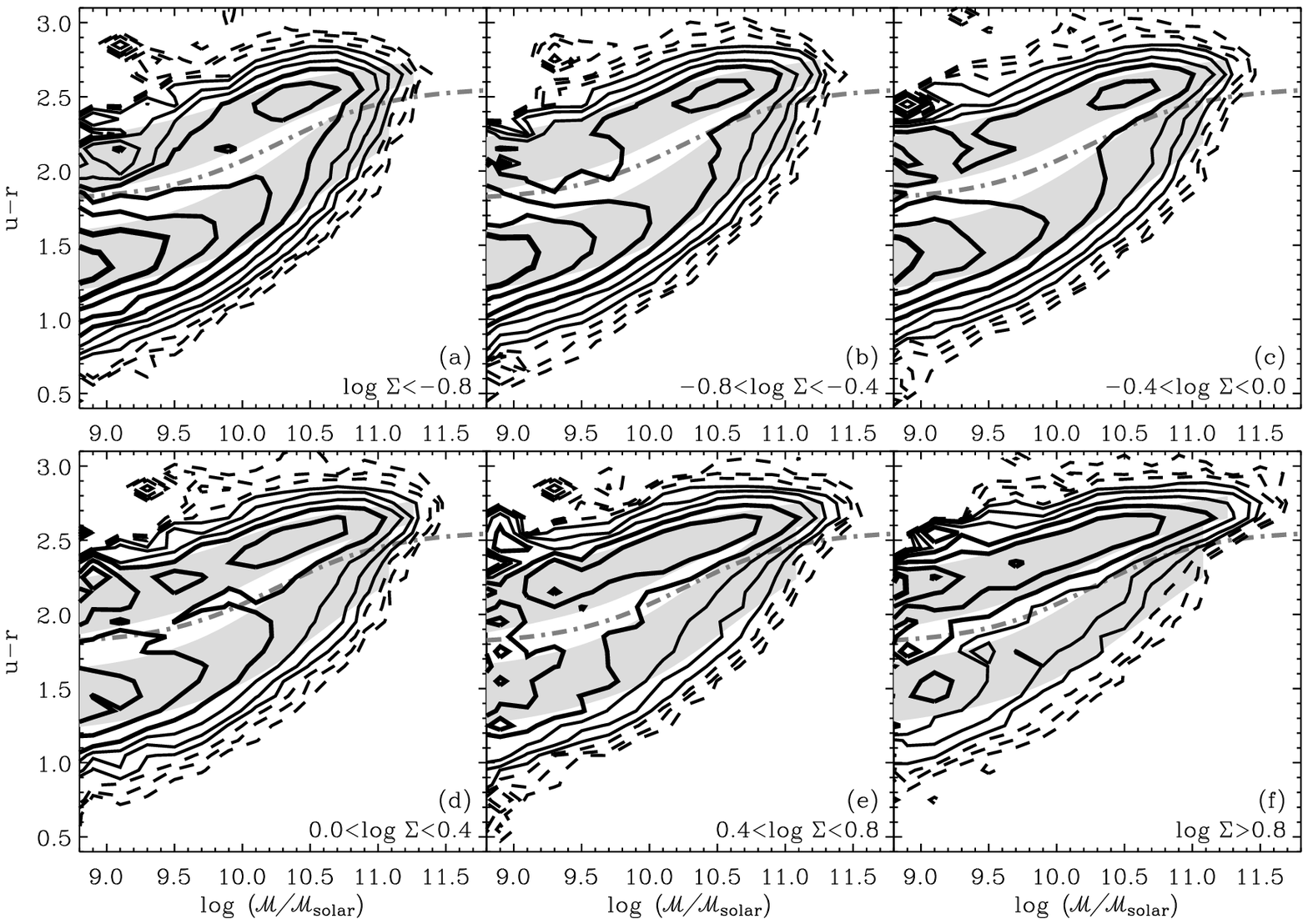}
\caption{Colour versus stellar mass relations for different environments.
  Panel~(a) represents void-like environments while panel~(f) represents
  cluster-like environments. The dashed and solid lines represent
  logarithmically-spaced density contours. The grey regions outline the red
  and blue sequences. The dash-and-dotted lines show the best-fit divider from
  Fig.~\ref{fig:compare-cm}(c).}
\label{fig:colour-mass}
\end{figure*}

\subsection{Environmental dependence of the galaxy stellar mass functions}
\label{sec:enviro-gsmf}

Before considering the variation of the colour-mass relations it is important
to note that the galaxy stellar mass functions (GSMFs) vary significantly with
environment. The galaxy population was divided into 12 environment bins and 16
mass bins.  
Figure~\ref{fig:mass-functions} shows the variation of the GSMFs with
environment. The GSMFs are plotted using mass fraction normalised by total
mass within each environment bin (i.e.\ the integral under each curve is
unity). This has two advantages: (i) the $y$-axes values range over 2 orders
of magnitude compared to about 4 when plotting number density; and (ii) the
peak of each GSMF corresponds to the dominant contribution to the total
stellar mass within an environment bin.  This varies from $\log \mass \sim
10.6$ to 11.0 from low to high densities; with the fractional contribution due
to high mass galaxies ($\log \mass \sim 11.5$) increasing by a factor of about
50.

The GSMFs can be characterised by fitting \citet{Schechter76} functions:
\begin{equation}
  \phi_{N}         \: \propto \: \mass^{\alpha}   e^{-\mass/\mass^{*}}
\label{eqn:schechter-proportional}
\end{equation}
where $\phi_N \, \dd \mass$ is the number of galaxies with masses 
between $\mass$ and $\mass + \dd \mass$; or
\begin{equation}
  \phi_{\log\mass} \: \propto \: \mass^{\alpha+2} e^{-\mass/\mass^{*}}
\label{eqn:schechter-log-mass}
\end{equation}
where $\phi_{\log\mass} \, \dd \log\mass$ is the total mass of galaxies
between $\log\mass$ and $\log\mass + \dd \log\mass$.  Normalised
$\phi_{\log\mass}$ is plotted in Fig.~\ref{fig:mass-functions}.  For a
standard Schechter function, the peak of the $\phi_{\log\mass}$ GSMF is given
by $\mass_{\rm peak} = \mass^{*} (\alpha+2)$. Thus, $\mass_{\rm peak}$ can be
considered an alternative characteristic mass for faint-end slopes around
$-1$, which is less dependent on model choice (cf.\ \citealt{Andreon04}).

The Schechter fits are shown in Fig.~\ref{fig:mass-functions}(b) with $\log
\mass^{*}$ versus environment plotted in Fig.~\ref{fig:mass-functions}(c).
A fit to $\log\mass_{\rm peak}$ is shown by the dotted line, formally
\begin{equation}
 \log [\mass^{*} (\alpha+2)] = (10.73 \pm 0.01) + (0.15 \pm 0.01) \log\Sigma
\label{eqn:m-star-fit}
\end{equation}
The faint-end slope is approximately $-1$ over the fitted range which does not
vary strongly with environment. A standard Schechter function does not fit the
entire mass range. Fits to the extended mass range using a double Schechter
function (Paper~I) are shown in the Appendix.

The increase of the characteristic mass with local density of 0.4\,dex,
equivalent to 1\,mag, is broadly consistent with other results.  A precise
quantitative comparison is difficult because of the different methods of
determining environment and different wavelengths.  \citet{croton05} found an
increase of $\sim1{\rm\,mag}$ between void and cluster-like environments using
$b_J$ luminosities.  \citet{hoyle05} found a difference of 0.9\,mag between
void and wall galaxies using $r$-band luminosity.  There are many other
analyses looking at luminosity functions versus group masses, cluster
properties, etc.\ (e.g.\ \citealt{BST88,balogh01,depropris03,ZMM06}).  The
main point is that the overall characteristic luminosity or mass increases
with environmental density. This should be accounted for when comparing galaxy
colours between different environments.

\begin{figure*}
  \includegraphics[width=\doublecolsize\textwidth]{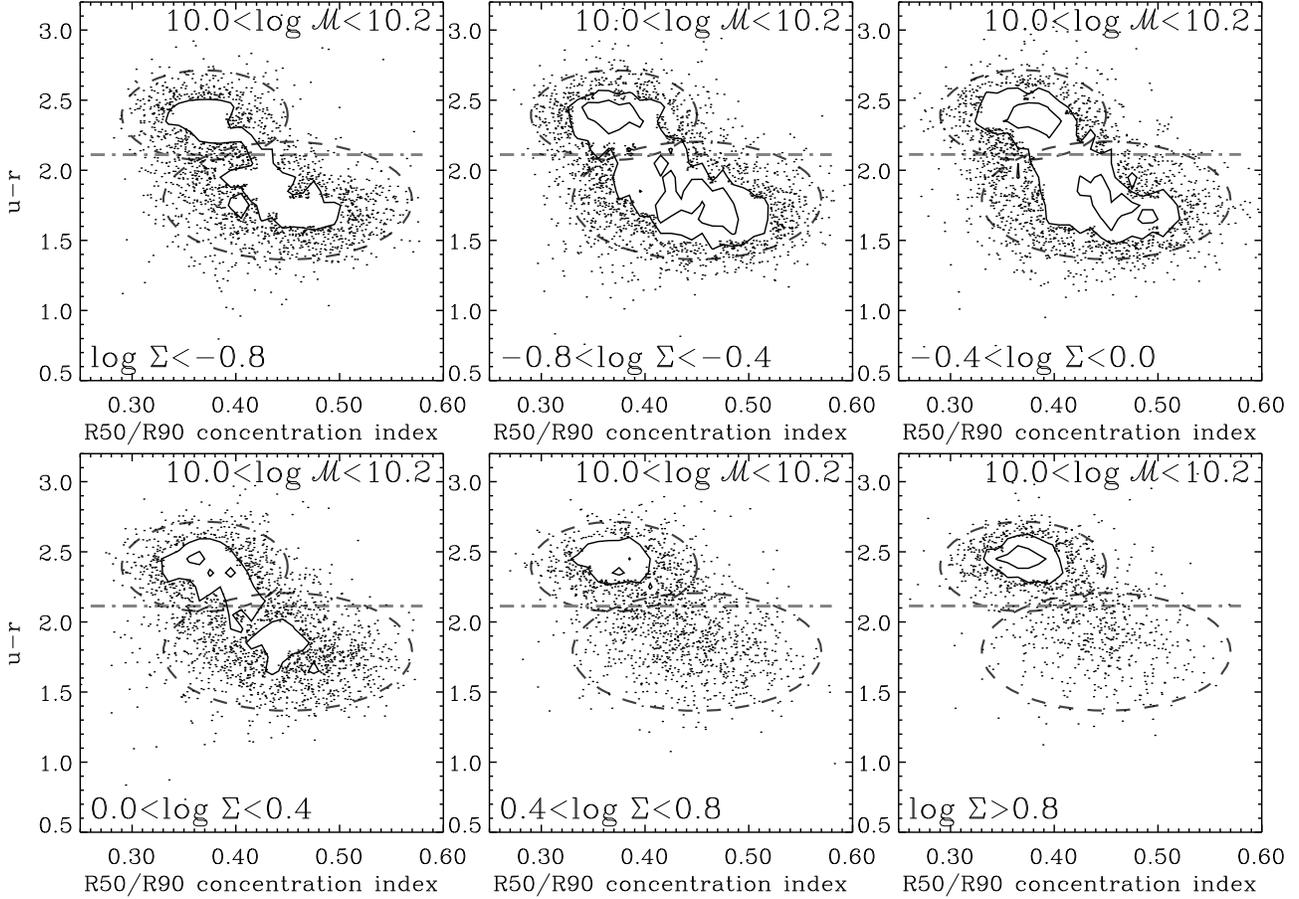}
\caption{Colour versus concentration index across different environments
  for stellar masses around $10^{10}\,\Msun$.  The points and solid contours
  represent galaxies in our sample.  The dashed ellipses are fixed in each
  panel outlining the locus of the red- and blue-sequence galaxies.  The
  dominant effect is clearly the increase in the fraction of galaxies on the
  red sequence with increasing density; while the locus of the sequences
  become marginally redder. The dash-and-dotted lines show the best-fit
  color divider for the mean mass.}
\label{fig:colour-conc-enviro}
\end{figure*}

\subsection{Environmental dependence of the colour-mass 
  and colour-concentration relations}
\label{sec:enviro-colour-mass-conc}

The galaxy population was divided into environmental bins in order to
determine the variation of the colour-mass relation.
Figure~\ref{fig:colour-mass} shows the colour-mass relations for six
environmental bins ranging from $\log\Sigma<-0.8$ to $\log\Sigma>0.8$. The
equivalent of the cluster red sequence is evident in
Fig.~\ref{fig:colour-mass}(f); while the dominance of the blue sequence at
lower masses is evident in Fig.~\ref{fig:colour-mass}(a-c) for the lower
densities.  While the red sequence is dominant at higher masses in all
environments, the upper mass cutoff is increasing with environmental density.
The best-fit dividing line from \S\,\ref{sec:colour-mag-mass} provides a
reasonable color division between the two sequences in all environments as
shown by the dash-and-dotted lines.

As noted in Paper~III, the mean positions of the sequences as
shown by the grey regions in Fig.~\ref{fig:colour-mass} do not vary strongly
with environment. This is also evident if one plots colour versus concentration
index as a function of environment for different mass ranges.
Figure~\ref{fig:colour-conc-enviro} shows this for galaxies with masses around
$10^{10}\,\Msun$.  The clearly dominant effect is the increase in the fraction
of galaxies on the red sequence as a function of environment. 

\subsection{Environmental and stellar mass dependence 
  of the relative red- and blue-sequence numbers}
\label{sec:enviro-mass-unify}

\begin{figure*}
  \includegraphics[width=\doublecolsize\textwidth]{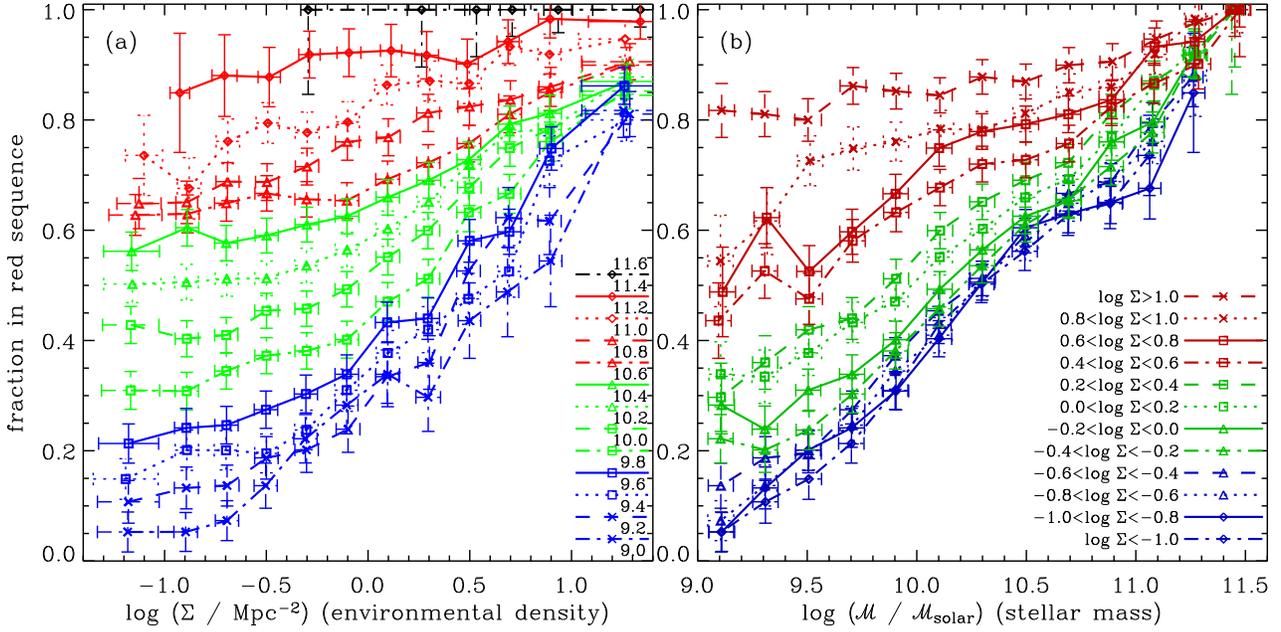}
\caption{Fraction of red-sequence galaxies 
  versus environment and versus stellar mass.  In panel (a), the symbols and
  lines represent different stellar masses as shown in the legend ($\log\mass$
  from 9.0 to 11.6). In panel (b), the lines represent different environmental
  densities.  Systematic errors of 0.03 were added in quadrature to the
  Poisson errors.  Note the similarity between the two plots leads to the
  unification schemes shown in Fig.~\ref{fig:frac-red-adj}.}
\label{fig:frac-red}
\end{figure*}

In \citet{baldry04conf}, it was shown that the fraction of galaxies in the red
sequence could be unified by a function of environmental density and
luminosity.  Here, we proceed with a similar analysis using stellar mass in
place of luminosity.  The population was divided into 12 environment bins and
13 mass bins.  The fraction is defined as galaxies redder than the best-fit
dividing line [Fig.~\ref{fig:compare-cm}(c) and~\ref{fig:colour-mass}].
Figure~\ref{fig:frac-red} shows the fraction of red-sequence galaxies (a)
versus environment for different stellar masses and (b) versus stellar mass
for different environments. Qualitatively the plots are similar: with
increasing red fraction versus stellar mass or environment and a greater
variation at the lower end of each scale.

\begin{figure*}
  \includegraphics[width=\doublecolsize\textwidth]{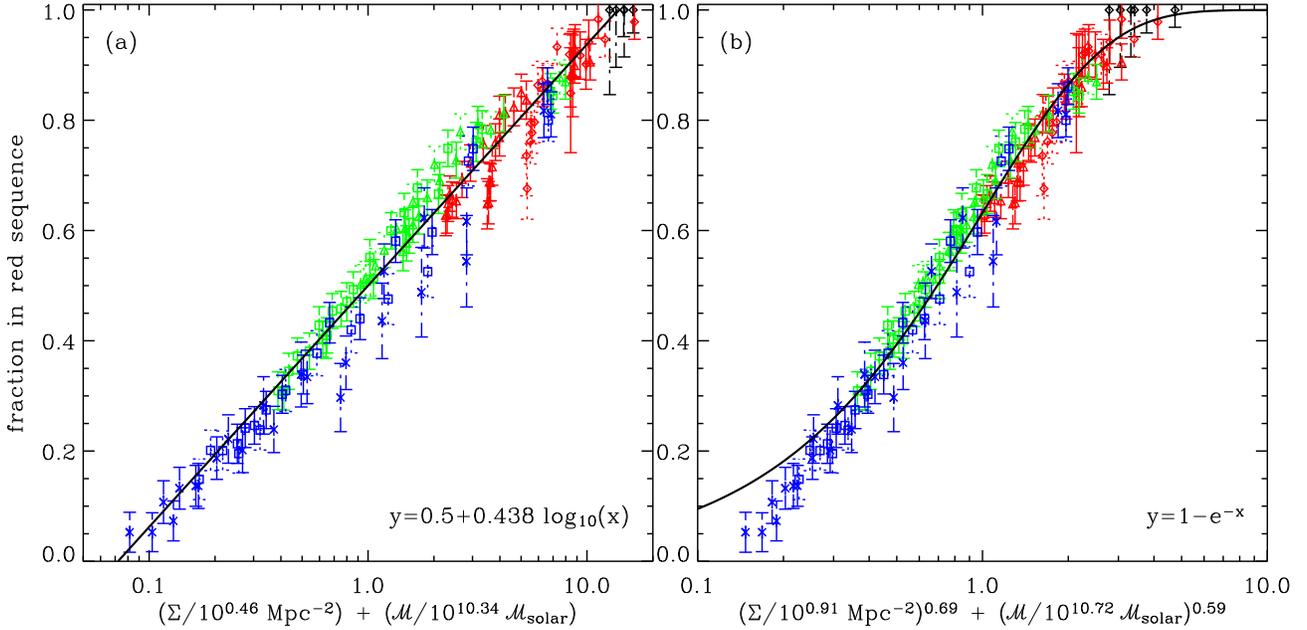}
\caption{Fraction of red-sequence galaxies versus unified quantities: 
  (a) using Eq.~\ref{eqn:fracA} and (b) using Eq.~\ref{eqn:fracB}.  The
  symbols and error bars represent different stellar masses as shown in
  Fig.~\ref{fig:frac-red}(a). The solid black lines represent the functions
  shown in the bottom-right corner of each panel.}
\label{fig:frac-red-adj}
\end{figure*}

Both stellar mass and environment affect the probability of a galaxy being in
the red sequence. First of all, we found an empirical relation that is based
on the sum of projected environmental density and stellar mass:
\begin{equation}
  f_r = 0.5 + a_1 \log_{10}( \Sigma / a_2  +  \mass / a_3 )
\label{eqn:fracA}
\end{equation}
(with the constraint $0 \le f_r \le 1$) where $f_r$ is the fraction of
red-sequence galaxies and $a_{1,2,3}$ are the parameters to determine.
Secondly, we found an empirical relation more naturally related to probability
theory that is given by:
\begin{equation}
  f_r = 1 - \exp\{ - [(\Sigma / b_1)^{b_2} + (\mass /b_3)^{b_4}]  \}
\label{eqn:fracB}
\end{equation}
where $b_{1,2,3,4}$ are the parameters to determine.  This function has the
useful feature of asymptotically tending to 0 and 1 at the extremes.
Figure~\ref{fig:frac-red-adj} shows these empirical relations that were
determined by minimising $\chi^2$. The values of the parameters are (a) 0.438,
$10^{0.46}\perMpcsq$, $10^{10.34}\Msun$ and (b) $10^{0.91}\perMpcsq$, 0.69,
$10^{10.72}\Msun$, 0.59.  [Note an empirical law of the form $f_r =
F(c\log\Sigma + \log\mass)$ does not work; the $\chi^2$ is about 200 higher
for a quadratic function $F$.]

Both empirical relations provide a reasonable fit to the data.  The relation
in Fig.~\ref{fig:frac-red-adj}(b) provides a marginally better fit relative to
Fig.~\ref{fig:frac-red-adj}(a) despite the deviation at low masses.  In
addition, the low mass end is uncertain because of the difficulty in extracting
a reliable low-redshift sample; there are systematic uncertainties mostly
related to deblending, for example, some low-redshift systems are in fact
deblends of large galaxies \citep{blanton05}.  For $\log\mass>9.4$, the
results are more reliable because the average redshifts are higher. If
$\chi^2$ is determined only using galaxies with $\log\mass>9.4$ then the
relation in Fig.~\ref{fig:frac-red-adj}(b) provides a significantly better
fit.  GSMFs for the red- and blue-sequences based on these empirical laws are
shown in the Appendix.

\clearpage

\section{Discussion}
\label{sec:discussion}

The advent of large-volume redshift surveys has greatly expanded the studies
of galaxy populations as a function of environment that had traditionally been
based around clusters \citep{Dressler80} to the full range from `void' to
cluster. There are many studies including Paper~II to this paper
(\S\,\ref{sec:earlier-work}) that focus on the variation of galaxy properties
with local density
\citep{kauffmann04,tanaka04,KFS05,alonso06,haines06,mateus06,patiri06}.  Other
studies have analysed the average density as a function of galaxy property
\citep{hogg03,blanton05enviro}, the clustering properties of galaxies
\citep{wild05,zehavi05,li06}, and the relationship between galaxies and dark
matter derived from galaxy-galaxy weak lensing \citep{gray04,mandelbaum06}.
The data available for $z<0.2$ is the most comprehensive, e.g., from SDSS and
2dFGRS data, but gains have been made at higher redshift
\citep{wilman05,yee05,cooper06,ilbert06}.  While local density measurements
illuminate various environmental trends in the data, galaxies in
semi-analytical models are primarily associated with dark-matter halos.  The
main focus of this discussion is comparing models with the data by `measuring'
the analogous quantity to projected galaxy density in the models.

Our results show that the local galaxy population divides neatly into two
types, and that the fraction of each type depends exclusively on a combination
of stellar mass and local environment.  Importantly, the environmental
dependence is not a second-order effect, but is at least as important as
stellar mass in determining the fraction of red galaxies in a population.  By
contrast, in simple galaxy formation models, the characteristics of a galaxy
are usually determined primarily by the mass of its dark matter halo, which is
closely related to the stellar mass of the galaxy.  That is because the
formation time, cooling rate and merging history of a halo are most strongly
related to its present day mass.  The most important environmental
consideration in most of these models is that galaxies at the centre of a dark
matter halo are treated differently from non-central, or satellite galaxies
\citep[e.g.][]{cole00}.  Only the central galaxies are assumed to possess a
halo of hot gas that can potentially cool and replenish the disk.  In
addition, there are second-order effects related to the large-scale density
field that, qualitatively at least, can mimic some of the observed trends with
environment \citep{maulbetsch06}, depending on how the details of star
formation and feedback are treated.

Our results lead naturally to two questions in this context.  One is, how does
our measurement of environment, $\Sigma$, relate to the dark matter density
field $\Delta\rho/\rho$?  And the other is whether or not galaxy formation
models that are successful in other respects are also able to reproduce the
observed dependence of galaxy colour on environment.  We will address these
questions using the $z=0$ output of the Virgo Consortium's Millennium
Simulation.  The details of the dark matter are described in
\citet{springel05nat}, and we will compare with the galaxy formation models of
both \citet{croton06} and \citet{bower06}.  These models are improved over
earlier efforts \citep[e.g.][]{cole00} in that, by including a model of
feedback from AGN, they are able to better match the observed colour
distribution as a function of galaxy luminosity.  Specifically, the inclusion
of AGN feedback removes most of the bright blue galaxies, and increases the
colour difference between the red and blue populations \citep[see
also][]{SDH05}.

Although both models include feedback from radio-jets and AGN, the models use
different schemes to implement the feedback. \citet{croton06} compute an
energy feedback rate based on a semi-empirical model involving the mass of the
host halo and that of the central black hole.  Their paper discusses how the
expression they use is related to the Bondi accretion rate. In contrast, the
\citet{bower06} model assumes that AGN feedback will be self-regulating if the
cooling time is long compared to the sound crossing time of the system so that
the cooling of gas takes place from a quasi-hydrostatic atmosphere
\citep{Binney04}. The model also requires that central black hole is
sufficiently massive to provide heating rate.  In addition to the treatment of
AGN jets, the models also differ in many details of the implementation of
cooling galaxy merging, star bursts and many other factors.

\subsection{Galaxy versus dark matter densities in the models}

To construct a mock-observational sample to compare with our data, we select
galaxies at $z=0$ from the simulation.  We restrict most of the analysis to
galaxies with stellar masses $\mass>10^{10}\Msun$, which typically belong to
dark matter halos with at least 100 particles (so their merger histories are
reasonably well resolved).  However, to define the local density we will
select a luminosity-limited sample (see below) that includes lower-mass
galaxies.  The redshift of each galaxy is determined as $z=\left(H_\circ
  r_z+v_z\right)/c$, where $r_z$ and $v_z$ are arbitrarily taken to be the
position and peculiar velocity in the $z$-axis-coordinate of the simulation.
The simulation box is 714\,Mpc on a side, large enough that we can select
the redshift range $0.01<z<0.085$ to match our data sample, without the need
to tile the box.\footnote{Distances, absolute magnitudes and masses in the
  simulation are converted to a cosmology with
  $H_0={\rm\,70\,km\,s^{-1}\,Mpc^{-1}}$. The SDSS $r$-band luminosity is
  computed from the $B$- and $V$-band magnitudes in the catalogue via the
  transformation $r=V-0.42(B-V)+0.21$.  This is based on the calibration of
  \citet{jester05}, with an additional 0.1\,mag shift to ensure the space
  density of galaxies with $M_r=-20$ matches that of the data.}

We measure $\Sigma$ in the simulation using a similar prescription to that
used for the data.  Specifically, for each galaxy we define a DDP as all
galaxies in the galaxy catalogue brighter than $M_r=-20$, and within $cz=1000$
km/s.  The distances to the fourth- and fifth-nearest neighbours within this
DDP are calculated for each galaxy in the simulation, and the projected
density $\Sigma$ is computed as for the data.  We do not attempt to model the
SDSS spectroscopic completeness, and thus ignore the small effect of only
having photometric redshift estimates in incomplete regions (galaxies with
large uncertainties are not included in the data analysis in any case).

\begin{figure}
  \includegraphics[width=\singlecolsize\textwidth]{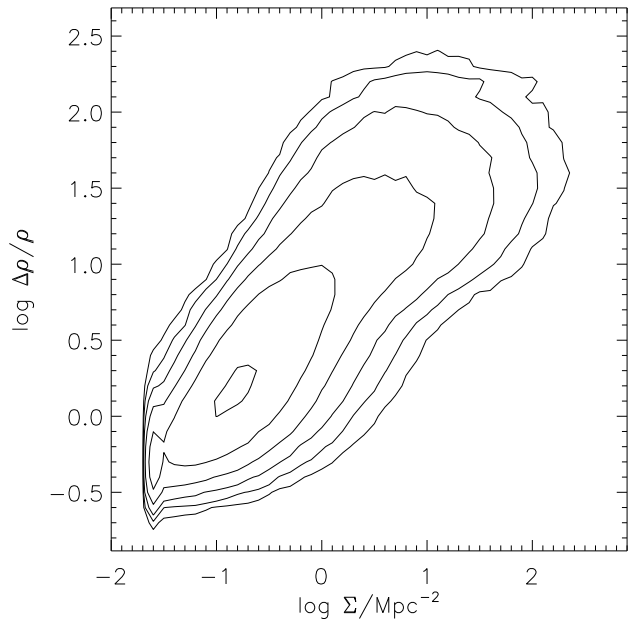}
\caption{The correlation between dark matter overdensity
  $\Delta\rho/\rho$, smoothed with a Gaussian kernel of radius $1.78$
  Mpc, and projected galaxy surface density $\Sigma$, from the model
  of \citet{croton06}.  The data are binned by 0.1 in the logarithm of
  each quantity, and the contours are logarithmically spaced by 0.4
  with the lowest contour representing 75 galaxies per bin.  The
  correlation demonstrates that the observationally defined $\Sigma$
  broadly traces the underlying dark matter density distribution.}
\label{fig:mill-density-full}
\end{figure}

\begin{figure}
  \includegraphics[width=\singlecolsize\textwidth]{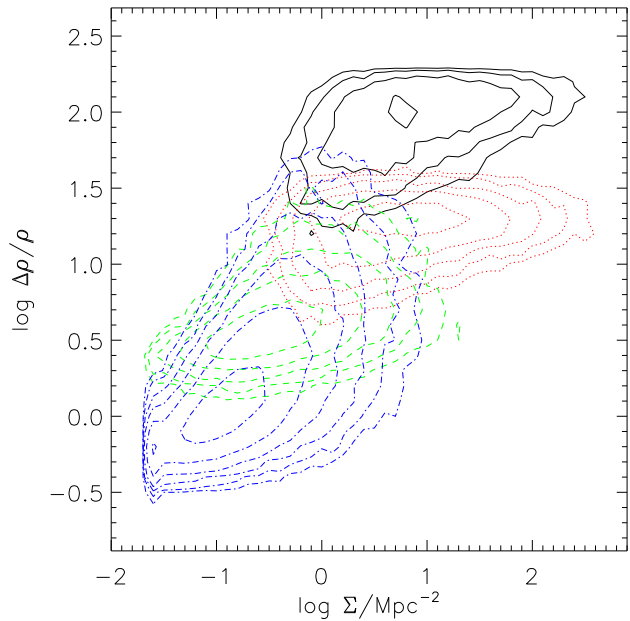}
\caption{As Fig.~\ref{fig:mill-density-full}, but for four distinct
  populations divided by the mass of the largest dark matter halo in which
  they are embedded.  The lowest contour level represents 10 galaxies per bin
  (0.1 in the logarithm of $\Sigma$ and $\Delta\rho/\rho$), and the contours
  are logarithmically spaced by 0.4.  The solid contours are restricted to
  galaxies with $\log{\mass_{halo}/\Msun}=14.75\pm0.1$; the dotted contours
  represent $\log\mass_{halo}=13.75$, the dashed contours
  $\log\mass_{halo}=12.75$, and the dash-and-dotted contours
  $\log\mass_{halo}=11.75$.  For halos of fixed mass, $\Sigma$ spans a wide
  range, as it is sensitive to the local environment of galaxies.}
\label{fig:mill-density}
\end{figure}

Figure~\ref{fig:mill-density-full} shows how the measurement of $\Sigma$
compares with the smoothed dark matter overdensity, $\Delta \rho/\rho$,
computed with a Gaussian kernel of radius 1.78\,Mpc, for all galaxies in the
\citet{croton06} catalogue with $M_r\leq-20$.  There is a clear correlation,
in that galaxies with higher $\Sigma$ tend to lie in regions that are
overdense in dark matter.  This is encouraging but not surprising.
Figure~\ref{fig:mill-density} shows how this correlation depends on halo mass,
by which we mean the mass of the largest dark matter halo associated with the
galaxy (i.e., not the subhalo in which the galaxy may reside).  The three sets
of contours correspond to halos with masses within 0.1 of
$\log(\mass/\Msun)=14.75$, $13.75$, $12.75$ and $11.75$.  For halos of a
given mass, there is a large range in $\Sigma$; thus, this estimate of local
density is sensitive to variations in environment within a given halo.
Moreover, the distribution of $\Sigma$ is only weakly dependent on mass, for
$\mass\ga 10^{13}\Msun$, which reflects the fact that the density structure of
dark matter halos are nearly self-similar.  On the other hand, the dark
matter overdensity on $1.78$ Mpc scales is more sensitive to halo mass,
because the size of the smoothing kernel is comparable to, or larger than, the
typical halo.  For low-mass halos, where the number of bright $M_r<-20$
galaxies per halo is small, $\Sigma$ traces dark matter overdensity reasonably
well, because both are probing similar scales.

\subsection{Comparison between models and data}

Recently, \citet{weinmann06} have demonstrated that although the models of
\citet{croton06} do a good job of reproducing the overall
luminosity-dependence of the observed galaxy colour distribution, they do not
succeed in matching the colour distribution in groups and clusters.
Specifically, they found that the model predicts too many faint, red galaxies
in these large halos.  We will investigate this further by characterising
environment using the continuous variable $\Sigma$, which is more closely
related to the large-scale dark matter density field than to the mass of the
halo.  

\begin{figure*}
  \includegraphics[width=\middlecolsize\textwidth]{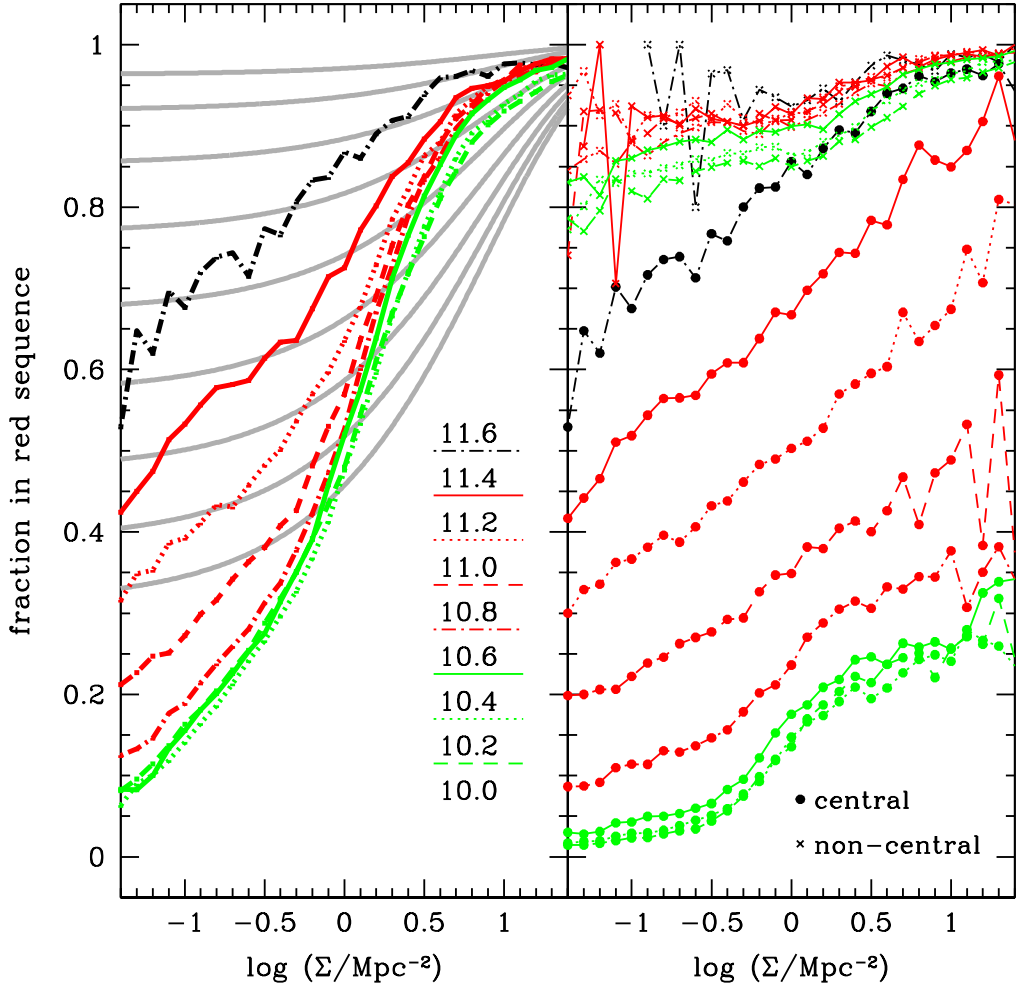}
\caption{{\bf Left panel:} The fraction of red galaxies 
  in the \citet{croton06} model, as defined in the text, as a function of
  stellar mass and environment.  The green, red and black lines represent the
  model. They are analogous to those shown in Fig.~\ref{fig:frac-red}(a) but
  only extend to masses as low as $10^{10}\Msun$.  For reference, the grey
  lines represent $f_r$ from the data parameterised using the unified relation
  of Fig.~\ref{fig:frac-red-adj}(b) for $\log\mass=10.0$ to 11.6.  {\bf Right
    panel:} The model galaxies are divided into central galaxies (solid
  points) and satellite galaxies (crosses).}
\label{fig:mill-fred}
\end{figure*}

To divide the model into red and blue galaxies we follow the approach of
\citet{weinmann06} and make a magnitude-dependent colour cut.  We find that a
separation at $(B-R)=-0.05\left(M_r+20\right)+1.83$ provides a good separation
between the red and blue populations of the model (our results below are not
sensitive to this cut, because the model populations are so well separated in
colour).  
Figure~\ref{fig:mill-fred} shows the fraction of red galaxies, using this
criterion, as a function of stellar mass and local density $\Sigma$.
Qualitatively, the models reproduce the observations well, as the red fraction
increases with both stellar mass and $\Sigma$.  However, quantitatively there
are some interesting differences.  In particular, the dependence in the model
is actually too strong relative to the data.  At low densities, the model
predicts too few red galaxies, with a red fraction that varies from 0.05 to
0.6 over the plotted mass range.  By contrast, over the same mass range, the
data show a red fraction ranging from 0.4 to 1.0.  Another point of interest
is that the mass dependence of the red fraction in the models is weak below
$\sim 10^{10.6}\Msun$, while the data show a continuous decrease in red
fraction with stellar mass over another order of magnitude.  At the
high-density end, the models seem to do a better job of reproducing the high
red fraction and weak stellar-mass dependence seen in the observations.

The main source of the environmental dependence in the model is simply
illustrated in the right panel of Fig.~\ref{fig:mill-fred}, where we divide
the sample into central and satellite galaxies.  As shown by
\citet{weinmann06}, most satellite galaxies are red, independent of luminosity
and density, in the models.  The central galaxies, on the other hand, have a
strong stellar mass dependence, and a dependence on environment that is
considerably weaker than that shown by the population as a whole, in the left
panel.  The overall environmental trend is therefore largely due to the fact
that low-density regions are dominated by central galaxies, while the
high-density end is dominated by satellite galaxies.  The main source of the
discrepancy with the observations, therefore, can be identified as the low
fraction of red galaxies amongst the central galaxies in low-density
environments.

\begin{figure*}
  \includegraphics[width=\middlecolsize\textwidth]{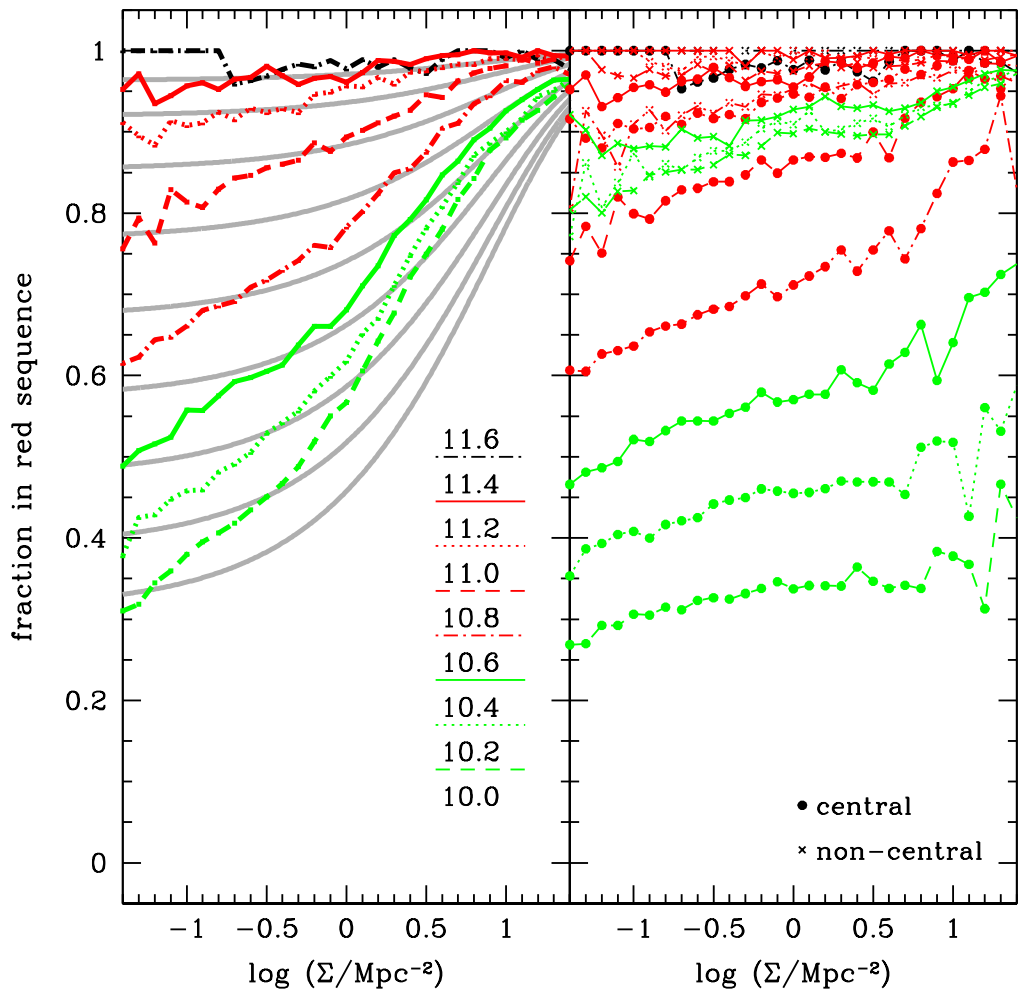}
\caption{As Fig.~\ref{fig:mill-fred}, but with the galaxy formation
  model of \citet{bower06}.  The main difference is that the central galaxies
  have a higher red fraction in all environments, leading to better agreement
  with the observed, overall environmental trend.  The model parameters were
  {\it not} adjusted to match these data.}
\label{fig:mill-fred-bower}
\end{figure*}

Figure~\ref{fig:mill-fred-bower} shows the analog of Fig.~\ref{fig:mill-fred}
for the \citet{bower06} models.  In these models, the colour-luminosity
relation is significantly different from \citet{croton06}, so we have had to
adopt a different definition of red galaxy, with a colour cut given by
$(B-R)=-0.07\left(M_r+20\right)+1.25$.  For this model, the agreement with the
observations is remarkably good.  The difference lies in the colour
distribution of the central galaxies, for which, at all densities, a larger
fraction lie in the red sequence than in the \citeauthor{croton06}\ model.
This is not surprising, as the main difference between the two models lies in
the treatment of AGN feedback, within central galaxies.  Therefore, the
satellite populations are largely unaffected.

\subsection{Implications}

\citet{weinmann06} showed that the \citet{croton06} model fails to predict the
presence of faint blue galaxies in groups and clusters, and link this
discrepancy to the uniformly red satellite population that is seen here in
both the \citeauthor{croton06}\ and \citet{bower06} model.  This is quite a
different effect from the discrepancy revealed by our results, as our data
show that the fraction of red galaxies in the densest environments is nearly
as high as the models predict.  Instead, our analysis shows a problem with the
{\it opposite} population in the \citeauthor{croton06}\ model, in that the
central galaxies in low-density environments are too frequently blue.  
Interestingly, this problem does not appear to be as severe in the
\citeauthor{bower06}\ model.  However, it is not clear whether this
improvement is due to their different prescription for AGN feedback, or
to some other aspect of the model.  Unambiguously identifying the root
cause of this difference requires a more detailed comparison between
the two models, which is beyond the scope of this paper.

The comparison between models and data that we have discussed and shown in
Figs.~\ref{fig:mill-fred}--\ref{fig:mill-fred-bower} is only the red-blue
sequence galaxy fraction; clearly colour-mass, colour-concentration relations
and GSMFs could also be compared as a function of environment. However,
colour-mass and colour-concentration relations are more dependent on the
detailed physics of stellar and galaxy evolution such as chemical and
dynamical evolution; whereas the red fraction (based on a colour cut dividing
the bimodal population) represents the the most significant variation with
environment.  The focus of this paper is on the unified relation of the red
fraction versus stellar mass and environment, and we have shown that
semi-analytical models with their distinction between central and satellite
galaxies can match this with an appropriate feedback description
(Fig.~\ref{fig:mill-fred-bower}). Of course this does not mean that the
\citet{bower06} model has the right answer for the galaxy bimodality as there
are many other pieces of data to match, e.g., lower mass galaxies, colour
relations and GSMFs. 

Given that the \citet{bower06} model produces a good match to the unified
relation we are now in a position to interpret it in the context of the model.
In the model, the likelihood that a galaxy lies on the red sequence
depends on two factors.  Firstly, the probability that the AGN feedback is able
to prevent the collapse of further gas depends on the mass and dynamical time
of the halo, and the mass of the central black hole. Both of these factors
become more effective in larger mass galaxies (through the correlation between
central galaxy mass and halo mass, and through the correlation between bulge
mass and black hole mass, respectively). Secondly, the probability that a
galaxy is a satellite (and therefore unable to accumulate cold gas) increases
as $\Sigma$ increases.  As the galaxy mass increases, the galaxy is more
likely to be central and red regardless of the $\Sigma$ parameter.  Conversely
the galaxies with the lowest mass are very unlikely to be central if $\Sigma$
is high. Thus the sensitivity to $\Sigma$ depends strongly on galaxy mass, and
this dependence is reflected in the unified relation that we have presented.

The success of the model in representing the impact of environment may at
first seem surprising since the models presented do not yet include many
environment-related process such as ram pressure stripping of cold disk gas
and tidal interactions between satellite galaxies. In the models, interaction
with the environment occurs because the galaxies external gas reservoir is
removed as it becomes a satellite galaxy orbiting within a larger halo.
However, because the models assume a rapid interchange between the cold disk
gas and the external gas reservoir, the loss of the external halo quickly
suppresses the formation of disk stars. It seems that this provides a good
approximation to the effect of the complete suite of environmental
interactions. The caveat, of course, is that the process (as currently model)
seems overly effective so that too few satellite galaxies are able to continue
star formation, even in if the parent halo has relatively low mass. Clearly
this is an important area in which the models can be developed by introducing
a more complete description for the removal of the external reservoir.

\section{Summary}
\label{sec:summary}

We have analysed a primary sample of 151\,642 galaxies from the SDSS
determining stellar masses, rest-frame colours and projected environmental
densities.  The peak, or characteristic mass, of the GSMFs increase by about
0.4\,dex from $\log\Sigma\simeq-1.2$ to $\log\Sigma\simeq+1.3$
(Fig.~\ref{fig:mass-functions}). The galaxy population is bimodal with a red
and blue sequence for which the colour-mass and colour-concentration index
relations do not depend strongly on environment
(Figs.~\ref{fig:colour-mass}--\ref{fig:colour-conc-enviro}). In contrast, the
fraction of galaxies on the red sequence depends strongly on both stellar mass
and environment (Fig.~\ref{fig:frac-red}).  The stellar-mass and
environmental dependence of the red fraction can be unified using empirical
laws of the form given by Eqs.~\ref{eqn:fracA}--\ref{eqn:fracB}
(Fig.~\ref{fig:frac-red-adj}).

In order to compare models with the data, we computed the equivalent of galaxy
projected neighbour density for semi-analytical models based on the Millennium
Simulation. This showed that $\Sigma$ broadly traced the underlying
dark-matter density distribution
(Figs.~\ref{fig:mill-density-full}--\ref{fig:mill-density}).  The fraction of
red-sequence galaxies versus stellar mass and environment were computed for
two models \citep{bower06,croton06} and compared with the unified relation
(Figs.~\ref{fig:mill-fred}--\ref{fig:mill-fred-bower}).  Both models produce
qualitatively the right trends with environment; with the \citeauthor{bower06}
model providing a good quantitative match.

\section*{Acknowledgements}

We acknowledge NASA's Astrophysics Data System Bibliographic Services, the IDL
Astronomy User's Library, and IDL code maintained by D.~Schlegel (idlutils)
and M.~Blanton (kcorrect) as valuable resources.  We thank the Virgo
Consortium for allowing us access to the Millennium Simulation output.

Funding for the creation and distribution of the SDSS Archive has been
provided by the Alfred P.\ Sloan Foundation, the Participating Institutions,
the National Aeronautics and Space Administration, the National Science
Foundation, the U.S.\ Department of Energy, the Japanese Monbukagakusho, and
the Max Planck Society.

\appendix

\section{More on the galaxy stellar mass functions}
\label{sec:enviro-gsmf-sequences}

Galaxy luminosity functions and GSMFs are typically fitted using Schechter
functions; including galaxy type-dependent functions.  There is a logical
inconsistency here: the sum of Schechter functions do not sum to a Schechter
function in general. The main results in this paper suggest another method of
defining type-dependent functions. An overall GSMF is defined by a
Schechter-based function while the red- and blue-sequence functions are
defined using a probability or fractional function (e.g.\ Eq.~\ref{eqn:fracA}
or~\ref{eqn:fracB}).

Another key point is that an overall luminosity or mass function is in many
cases not well parameterised by a standard Schechter function. Modifications
to the Schechter function include: adding $\phi_{res}= 10^{(a + b M)}$
\citep{madgwick03}; adding a Gaussian function \citep{mercurio06}; and 
using a double Schechter function (Paper~I) given by
\begin{equation}
  \phi_X \, \dd X = e^{-X/X^{*}} 
  \left[ \phi^{*}_1 \left( \frac{X}{X^{*}} \right)^{\alpha_1} + 
         \phi^{*}_2 \left( \frac{X}{X^{*}} \right)^{\alpha_2} \right]
  \, \frac{ \dd X }{ X^{*} }
\label{eqn:double-schechter}
\end{equation}
where $\phi_X \, \dd X$ is the number density of galaxies with luminosity or
mass between $X$ and $X + \dd X$.  One can also specify $\alpha_2 < \alpha_1$
so that the second term will dominate at the faintest magnitudes (cf.\ 
\citealt{blanton05}). The first term then acts to increase the number density
around $X^{*}$.

\begin{figure*}
  \includegraphics[width=\doublecolsize\textwidth]{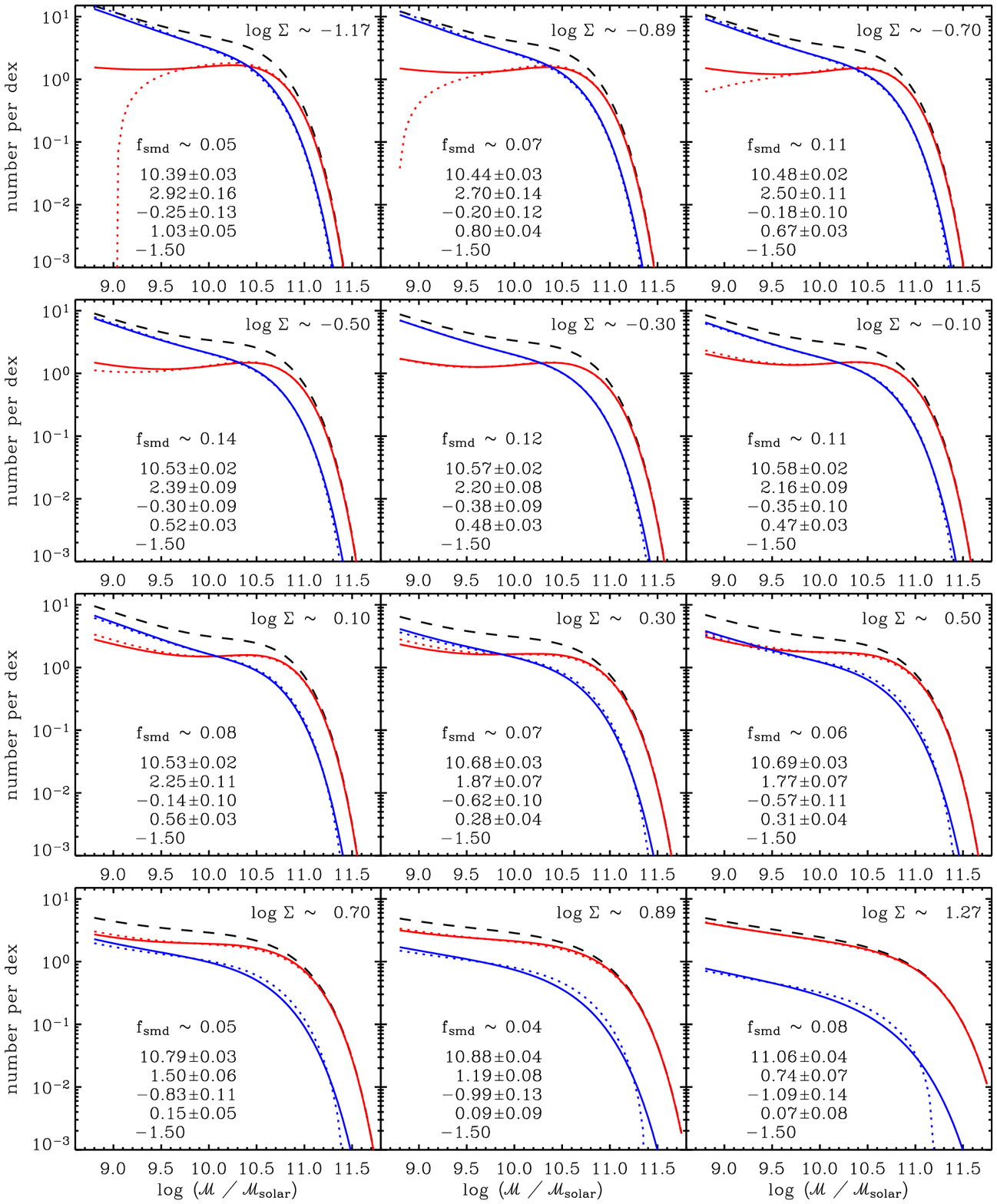}
\caption{Galaxy stellar mass functions.
  The dashed lines represent the double Schechter fits to the overall GSMFs.
  The solid lines represent the red- and blue-sequence GSMFs defined by
  multiplying the overall fit by $f_r$ and $1-f_r$ from Eq.~\ref{eqn:fracB};
  while the dotted lines represent the sequences using Eq.~\ref{eqn:fracA}.
  Each overall GSMF is normalised to a total mass of $10^{11}\,\Msun$.  See
  \S\,\ref{sec:enviro-gsmf}, Fig.~\ref{fig:mass-functions} and text of
  Appendix for further details.}
\label{fig:mass-functions2}
\end{figure*}

The standard Schechter fits are shown in Fig.~\ref{fig:mass-functions}(b) over
a reduced mass range. In order to fit the mass functions down to lower masses,
we used double Schechter functions with fixed $\alpha_2 = -1.5$.  The fixed
faint-end slope was the weighted best-fit average over all environments and
was consistent with \citet{blanton05}'s analysis of low-luminosity galaxies.
Figure~\ref{fig:mass-functions2} shows the fitted GSMFs for different
environments. The red- and blue-sequence GSMFs were then determined by
multiplying each overall GSMF by $f_r (\Sigma,\mass)$ and $1-f_r$ from
Eq.~\ref{eqn:fracA} or~\ref{eqn:fracB} with the parameters shown in
Fig.~\ref{fig:frac-red-adj}. The average $\log \Sigma$ value for each
environment, the approximate fractional contribution to the stellar mass
density ($f_{\rm smd}$) and the double Schechter parameters ($\log \mass^{*},
\phi^{*}_1, \alpha_1, \phi^{*}_2, \alpha_2$) are shown in each panel.  Note
with the IMF and mass-to-light ratios used in this paper, $\Omega_{\rm
  stars}\approx0.002$ (cf.\ \citealt{PS92,BG03,nagamine06}).


\label{lastpage}

\end{document}